\documentclass[12pt]{article}
\usepackage{amsmath}
\usepackage{graphicx}
\usepackage{natbib}
\usepackage{bbm}
\usepackage{url} 
\usepackage{xcolor}
\usepackage{amsfonts} 
\usepackage[hidelinks,colorlinks=true,linkcolor=blue,citecolor=blue]{hyperref}
\usepackage{algorithm}
\usepackage{algpseudocode}

\newcommand{\blind}{0}

\begin{document}

	\def\spacingset#1{\renewcommand{\baselinestretch}%
		{#1}\small\normalsize} \spacingset{1}

	
	\if0\blind
	{
		\title{\bf  Hysteretic Multivariate Bayesian Structural GARCH Model with Soft Information}
		\author{Tzu-Hsin Chien\\
			Graduate Institute of Statistics, National Central University, \\Taoyuan 320, Taiwan\\
			\\
			Ning Ning \\
			Department of Statistics, Texas A\&M University, College Station, \\TX 77843, USA\\
			\\
			and \\ \\
			Shih-Feng Huang\textsuperscript{*}\\
			Graduate Institute of Statistics, National Central University, \\Taoyuan 320, Taiwan}
		\date{}
		\maketitle
	} \fi

	\bigskip
	\begin{abstract}
		This study introduces the SH-MBS-GARCH model, a hysteretic multivariate Bayesian structural GARCH framework that integrates hard and soft information to capture the joint dynamics of multiple financial time series, incorporating hysteretic effects and addressing conditional heteroscedasticity through GARCH components. Various model specifications could utilize soft information to define the regime indicator in distinct ways. We propose a flexible, straightforward method for embedding soft information into the regime component, applicable across all SH-MBS-GARCH model variants. We further propose a generally applicable Bayesian estimation approach that combines adaptive MCMC, spike-and-slab regression, and a simulation smoother, ensuring accurate parameter estimation, validated through extensive simulations. Empirical analysis of the Dow Jones Industrial Average, NASDAQ Composite, and PHLX Semiconductor indices from January 2016 to December 2020 demonstrates that the SH-MBS-GARCH model outperforms competing models in fitting and prediction accuracy, effectively capturing regime-switching dynamics.
	\end{abstract}
	
	\noindent%
	\vfill
	
	
	\newpage
	\spacingset{1.75} 
	\section{Introduction}
	\label{sec:intro}
	
	Thanks to the rapid advancement of global communications, market practitioners now find it significantly more convenient to construct global investment portfolios and implement risk management strategies using real-time information, including both structured and unstructured data. However, this new environment also introduces a host of challenges. In particular, two primary types of information are typically available from global markets: hard (structured) and soft (unstructured) information. Hard information refers to numerical data, such as financial indicators and historical prices. In contrast, soft information encompasses unstructured textual content from online sources, such as market opinions, analyst commentary, and investor sentiment \citep{P2004}. Traditional statistical methods predominantly utilize hard information to model and forecast stock price dynamics. In the last decade, many machine learning techniques have been developed to analyze unstructured data and extract insights from soft information in sentiment classification, opinion recognition, and textual analysis (see \cite{TW2017, Jetal2019}, and references therein).
	
	Empirical studies have shown that financial time series often exhibit stylized features such as asymmetry, skewness, heavy tails, and volatility clustering \citep{T2010, HG2014, Hetal2021, HW2023, Lai2025}. Nonlinear time series models, including threshold autoregressive (TAR), ARCH, GARCH, and buffered AR models, have been widely employed to capture these characteristics \citep{TL1980, E1982, B1986, Zetal2014}. However, threshold models tend to perform poorly near regime boundaries, where abrupt changes occur \citep{WC2007}. To address this limitation, \cite{Letal2015} introduced the hysteretic autoregressive (HAR) model, incorporating hysteresis (a phenomenon widely observed in fields such as economics, engineering, and materials science \citep{G2002}) to smooth transitions between regimes. This modification mitigates the sudden regime-switching effects inherent in classical TAR models. Building on this idea, \cite{CT2016} extended the HAR model to a HAR-GARCH framework to better capture the volatility dynamics of financial data.
	
	Further developments include the multivariate Bayesian structural (MBS) model proposed by \cite{Qetal2018}, which characterizes the joint dynamics of multiple correlated time series using a rich set of components: time-varying trends, seasonal effects, cyclical patterns, and dynamic regression terms. The MBS model is particularly suitable for modeling non-stationary multivariate time series. Motivated by these observations, this study introduces a novel class of models, termed the SH-MBS-GARCH model, where `S' signifies the inclusion of soft information and `H' denotes the hysteretic mechanism. This model integrates both hard and soft information to capture the joint dynamics of multiple financial time series, incorporating hysteretic effects within the MBS framework and addressing conditional heteroscedasticity via GARCH components.

	A commonly used approach for drawing inferences from soft information involves transforming unstructured text data into structured form by computing Term Frequency–Inverse Document Frequency (TF-IDF) vectors from the collected texts \citep{M2004, Ketal2009, TW2017}.  However, directly performing classification using the complete set of words from daily news articles is typically infeasible due to the high dimensionality of the word list. To address this issue, dimension-reduction techniques are necessary to identify and retain the most informative keywords. \cite{Jetal2019} proposed a lexicon-based approach for sentiment classification, categorizing news articles by polarity (positive, negative, neutral) and Twitter posts by emotional tone (anxiety, calmness, dislike, fear, liking, love, joy, sadness, unknown). One key advantage of this lexicon-based method is that it avoids the labor-intensive task of labeling training data. \cite{Jetal2019} conducted sentiment analysis on both polarity and emotion indicators, incorporating the resulting sentiment predictors into the MBS model. Their empirical findings demonstrated that ARIMA, RNN, LSTM, and MBS models augmented with sentiment predictors outperform their counterparts without such predictors. Moreover, polarity-based sentiment predictors were found to have a significant influence on stock prices, whereas emotion-based predictors had little impact. Among all models considered, the MBS model with sentiment predictors achieved the best overall performance. 
	
	The SH-MBS-GARCH model class integrates hard and soft information to capture the joint dynamics of multiple financial time series, incorporating hysteretic effects within the MBS framework and addressing conditional heteroscedasticity through GARCH components. Various model specifications could utilize soft information to define the regime indicator in distinct ways. We propose a flexible, straightforward method for embedding soft information into the regime component, applicable across all SH-MBS-GARCH model variants. A Bayesian estimation framework, combining adaptive MCMC, spike-and-slab regression, and a simulation smoother, is developed to estimate model parameters. Extensive simulation and empirical analyses validate this approach. Simulation results confirm the algorithm’s ability to produce accurate parameter estimates. For empirical analysis, we utilize daily values of the Dow Jones Industrial Average (DJIA), NASDAQ Composite Index, and PHLX Semiconductor Index, alongside daily financial news from The New York Times, from January 2016 to December 2020. The numerical results consistently demonstrate that the SH-MBS-GARCH model effectively captures the regime-switching dynamics of joint financial time series.

	This paper is structured as follows. Section \ref{sec:Background_and_Motivation} provides the background and motivation for the SH-MBS-GARCH model. Section \ref{model} presents the proposed SH-MBS-GARCH model. Section \ref{sec:Parameter} outlines the Bayesian algorithm for parameter estimation, with prior distributions detailed in Section \ref{sec:Prior} and posterior distributions described in Section \ref{sec4.3}. Section \ref{sec:Numerical} reports simulation results and empirical studies, while Section \ref{sec:conclusion} summarizes findings and suggests directions for future research. Supplementary details on the proposed algorithm are included in the Appendix, with brief overviews of adaptive MCMC in \ref{secA.1}, spike-and-slab regression in \ref{secA.2}, and the simulation smoother method in \ref{secA.4}.
	
	\section{Background and Motivation for the SH-MBS-GARCH Model}
	\label{sec:Background_and_Motivation}
	
	In this section, we briefly introduce the structure of the MBS model in \cite{Qetal2018} (Section \ref{sec:MBSTS}), the HAR model in \cite{Letal2015} (Section \ref{sec:Hysteretic}), and the methodology for incorporating soft information in \cite{Jetal2019} (Section \ref{sec:Soft}), while outlining the motivation and necessity for proposing the SH-MBS-GARCH model.

	\subsection{Multivariate Bayesian structural models}
	\label{sec:MBSTS}
	
	Let $\boldsymbol{a}_t=(a_{t,1},\ldots,a_{t,m})^\top$ denote an $m\times1$ vector at time $t$. Specifically, an MBS model for an $m$-dimensional time series is defined as
	\begin{equation}
		\boldsymbol{y}_t=\boldsymbol{\mu}_t+\boldsymbol{\kappa}_t+\boldsymbol{\omega}_t+\boldsymbol{\xi}_t+\boldsymbol{\varepsilon}_t,
		\label{MBS}	
	\end{equation}
	where $\boldsymbol{\mu}_t$, $\boldsymbol{\kappa}_t$, $\boldsymbol{\omega}_t$, and $\boldsymbol{\xi}_t$ respectively denote the $m$-dimensional trend, seasonal, cyclical, and regression components of $\boldsymbol{y}_t$. Here, $\boldsymbol{\varepsilon}_t$ are independent and identically distributed (i.i.d.) as $N_m(0, \Sigma_\varepsilon)$, for $t = 1, 2, \ldots$. To simplify the notation, let $\boldsymbol{y}=(\boldsymbol{y}_1,\boldsymbol{y}_2,\ldots)$, $\boldsymbol{\mu}=(\boldsymbol{\mu}_1,\boldsymbol{\mu}_2,\ldots)$, $\boldsymbol{\kappa}=(\boldsymbol{\kappa}_1,\boldsymbol{\kappa}_2,\ldots)$, $\boldsymbol{\omega}=(\boldsymbol{\omega}_1,\boldsymbol{\omega}_2,\ldots)$, and $\boldsymbol{\varepsilon}=(\boldsymbol{\varepsilon}_1,\boldsymbol{\varepsilon}_2,\ldots)$.
	
	In \cite{Qetal2018}, the $m$-dimensional trend, seasonal, cyclical, and regression components of $\boldsymbol{y}_t$ in equation (\ref{MBS}) are defined as follows. For the time-varying local trend $\boldsymbol{\mu}_t$, let
	\begin{eqnarray*}
		\boldsymbol{\mu}_{t+1} &=& \boldsymbol{\mu}_t+\boldsymbol{\delta}_t+\boldsymbol{u}_t,\\
		\boldsymbol{\delta}_{t+1} &=& \boldsymbol{D}+\boldsymbol{\rho}(\boldsymbol{\delta}_t-\boldsymbol{D})+\boldsymbol{v}_t,
	\end{eqnarray*}
	where $\boldsymbol{u}_t \sim \mathcal{N}_m(0, \Sigma_u)$ and $\boldsymbol{v}_t \sim \mathcal{N}_m(0, \Sigma_v)$ are i.i.d. for $t = 1, 2, \ldots$, and are mutually independent. The term $\boldsymbol{\delta}_t$ represents the expected growth of $\boldsymbol{\mu}_t$ from time $t$ to $t+1$, $\boldsymbol{D}$ is an $m$-dimensional vector denoting the long-term slope, and $\boldsymbol{\rho} = \mathrm{diag}(\rho_1, \ldots, \rho_m)$ is an $m \times m$ diagonal matrix, where each $\rho_i \in (0, 1)$ specifies the local learning rate of the $i$th time series.

	For the seasonality, let the $i$th element of $\boldsymbol{\kappa}_{t}=(\kappa_{1,t},\ldots,\kappa_{m,t})^\top$ satisfies
	\begin{eqnarray*}
		\kappa_{i,t+1} &=& -\sum_{p=0}^{s_i-2}\kappa_{i,t-p}+w_{t,i},\qquad i=1,\ldots,m,
	\end{eqnarray*}
	where $s_i$ represents the number of seasons for the $i$th time series and $\boldsymbol{w}_{t}=(w_{1,t},\ldots,w_{m,t})^\top$, $t=1,2,\ldots$ are i.i.d. as $N_m(0,\Sigma_w)$.
	
	To capture dramatic short-term shocks to variations in a time series, a cyclical component $\boldsymbol{\omega}_t$ with a shock-damping parameter is further proposed by
	\begin{eqnarray*}
		\boldsymbol{\omega}_{t+1} &=& \boldsymbol{\zeta}\{\cos(\boldsymbol{\lambda})\boldsymbol{\omega}_t+\sin(\boldsymbol{\lambda})\boldsymbol{\omega}^*_t\}+\boldsymbol{\eta}_t\\
		\boldsymbol{\omega}^*_{t+1} &=& \boldsymbol{\zeta}\{-\sin(\boldsymbol{\lambda})\boldsymbol{\omega}_t+\cos(\boldsymbol{\lambda})\boldsymbol{\omega}^*_t\}+\boldsymbol{\eta}^*_t	
	\end{eqnarray*}	 
	where $\boldsymbol{\zeta}={\rm diag}(\zeta_{1},\ldots,\zeta_{m})$ and $\boldsymbol{\lambda}={\rm diag}(\lambda_{1},\ldots,\lambda_{m})$ are $m\times m$ diagonal matrices, and $\zeta_{i}\in(0,1)$ denotes the damping factor and $\lambda_{i}=2\pi/q_i$ with a cyclical period $q_i$ such that $\lambda_{i}\in[0,\pi]$ of the $i$th time series, $i=1,\ldots,m$. In addition, $\boldsymbol{\eta}_{t}$, $t=1,2,\ldots$ are i.i.d. $N_m(0,\Sigma_\eta)$, $\boldsymbol{\eta}^*_{t}$, $t=1,2,\ldots$ are i.i.d. $N_m(0,\Sigma_{\eta^*})$, and $\boldsymbol{\eta}_{t}$ and $\boldsymbol{\eta}^*_{t}$ are independent. In particular, when $\lambda_{i}=0$ or $\pi$, both $\boldsymbol{\omega}_{t}$ and $\boldsymbol{\omega}^*_{t}$ reduce to AR(1) processes. The main difference between the cyclical component and the seasonal component is the damping factor. The amplitude of the cyclical component decays in time, which is used to describe the effects of external shocks on a time series.
	
	For the regression factor $\boldsymbol{\xi}_t=(\xi_{1,t},\ldots,\xi_{m,t})^\top$, let $\xi_{i,t}$ satisfy
	\begin{equation}
		\xi_{i,t} = \boldsymbol{\beta}_i^\top \boldsymbol{z}_{i,t},
		\label{xi}
	\end{equation}
	where $\boldsymbol{z}_{i,t}$ is a $k_i$-dimensional vector of the covariates for the $i$th time series at time $t$ and $\boldsymbol{\beta}_i=(\beta_{i,1},\ldots,\beta_{i,k_i})^\top$ denotes regression coefficients, $i=1,\ldots,m$.
	
	\cite{Qetal2018} proposed a spike-and-slab method to estimate the parameters in equation (\ref{MBS}). One can apply the \textsf{R}-package \textsf{mbsts} to obtain the fitting results of the MBS model. The numerical results in \cite{Qetal2018} reveal that the MBS model outperforms a Bayesian structural model that treats each time series as independent, the ARIMA model with regression (ARIMAX), and the multivariate ARIMAX model (MARIMAX). This phenomenon indicates that one can improve the fitting and prediction performances by capturing the correlations among different time series. In addition, the MBS model can characterize the dynamics of multivariate correlated time series better than classical MARIMAX models. Based on these advantages, this study employs the MBS model as a crucial component in model building. 
	
	\subsection{Hysteretic time series models}
	\label{sec:Hysteretic}
	
	Let $r_t$ denote the log return at time $t$. \cite{Letal2015} proposed a HAR model to capture the widely observed hysteresis phenomena in economics and finance, aiming to characterize the nonlinear dynamics of time series. Specifically, a self-exciting HAR model, with regime indicator $R_t=j$ for $j\in \{0,1\}$, is defined by
	\begin{equation}
		r_t = 
		\phi^{(j)}_{0}+\displaystyle \sum_{i=1}^{p_j} {\phi^{(j)}_{i}r_{t-i}}+\sigma_j\epsilon_t,
		\label{HAR1}
	\end{equation}
	where $R_t$ is given by
	\begin{eqnarray}
		&&R_t = \left\{
		\begin{array}{ll}
			1  ,& \text{if } r_{t-d} \leq \tau_L,\\
			0  ,& \text{if } r_{t-d} > \tau_U,\\
			R_{t-1}  ,\quad & \text{otherwise.}
		\end{array}\right.
		\label{HAR2}
	\end{eqnarray}
	Here, $\epsilon_t$'s are i.i.d. random variables with zero mean and unit variance, $\sigma_{0}>0$ and $\sigma_{1} > 0$ are scalars, $d$ is a positive integer denoting the delay parameter, and $\tau_L \leq \tau_U$ are the boundary parameters of the hysteresis zone. When $\tau_L = \tau_U$, the model in equation (\ref{HAR2}) simplifies to the standard two-regime TAR model. The HAR model offers several advantages: it prevents abrupt transitions typically seen in traditional TAR models, improves model performance near regime boundaries, preserves the piecewise linear structure of TAR models, and provides a physically interpretable framework that is easy to understand.

	Along this line, \cite{CT2016} extended the HAR model to include further GARCH effects, denoted by HAR-GARCH herein. For example, a double HAR-GARCH model is represented by
	\begin{align}
		r_t = 
		\phi^{(j)}_0+\displaystyle \sum_{i=1}^{p_j} {\phi^{(j)}_{i}r_{t-i}}+\sqrt{h_t}\epsilon_t, 
		\label{HAR-GARCH1}
	\end{align}
	where $R_t=j$ for $j\in \{0,1\}$ is the regime indicator defined in equation (\ref{HAR2}) and
	\begin{align}
		\label{HAR-GARCH3}
		h_t =
		\alpha^{(j)}_{0}+\displaystyle \sum_{i=1}^{q_j} {\alpha^{(j)}_{i}a_{t-i}^2}+\displaystyle \sum_{i=1}^{m_j} {\beta^{(j)}_{i}h_{t-i}}. 
	\end{align}
	Here, $\{\epsilon_t\}$ is a white noise process with zero mean and unit variance, $\sqrt{h_t}=\sqrt{\text{Var}(r_t| \mathcal{F}_{t-1})}$ is the conditional volatility with $\mathcal{F}_{t-1}=\sigma(r_{t-1},r_{t-2},\ldots)$ being the information set up to time $t-1$, $p_1$ and $p_2$ are non-negative integers of the autoregressive orders, and $q_1$, $q_2$, $m_1$, and $m_2$ are the orders for volatility equations. Moreover, the coefficients $\{\phi^{(j)}_{i},i\ge0,j=0,1\}$ in equation (\ref{HAR-GARCH1}) and $\{\alpha^{(j)}_{i},\beta^{(j)}_{i},i\ge0,j=0,1\}$ in equation (\ref{HAR-GARCH3}) satisfy the stationary conditions of autoregressive and GARCH models, respectively. That is, $\sum_{i=1}^{p_j} {|\phi^{(j)}_{i}|}<1$ and $\sum_{i=1}^{q_j} \alpha^{(j)}_{i}+\sum_{i=1}^{m_j}{\beta^{(j)}_{i}}<1$ for $j=0,1.$ Chen and Truong (2016) proposed an MCMC algorithm to estimate the parameters in the HAR-GARCH model. Their empirical study finds strong evidence of the hysteretic effect.
	
	However, since $h_t$, $t = 1, 2, \ldots$, in equation (\ref{HAR-GARCH3}) is not constant in GARCH-type models, the process $R_t$ defined in equation (\ref{HAR2}) tends to cross the fixed threshold pair $(\tau_L, \tau_U)$ more frequently during periods of high volatility than during more stable periods.
	Such behavior is unrealistic in practice. For example, \cite{MM2000} observed that in bear markets, volatility typically increases with the duration of the downturn.
	As a result, using raw historical returns to define the regime indicator would lead to more frequent regime switches in bear markets compared to bull markets. To address this issue, we propose defining the regime indicator after filtering out GARCH effects, thereby enhancing its ability to accurately capture bull and bear market conditions. Additionally, we incorporate soft information into the regime indicator to reflect the influence of newly released economic news.

	\subsection{Soft information}
	\label{sec:Soft}
	
	To capture the influence of macroeconomic soft information, we employ the EPU Index in our analysis. The EPU Index has been widely utilized in various empirical studies across economics and finance.
	For example, \cite{Betal2020} applied the EPU data to assess near- and medium-term macroeconomic effects of the COVID-19-induced uncertainties. \cite{Yetal2021} adopted the EPU data to investigate its impact on carbon emission intensity at the firm level in China. \cite{Cetal2023} investigated the predictive ability of categorical EPU indices for stock-market returns. Specifically, we obtained the U.S. Daily News EPU Index from \texttt{http://www.policyuncertainty.com} and denote it by $D_{1,t}$.

	Furthermore, we incorporate the daily news reported in, for example, the New York Times, of each index. Inspired by \cite{Metal2014} and the references therein, \cite{Jetal2019} adopted the following five-step lexicon-based approach to draw soft information. In particular, they used SenticNet as the lexicon, which provides a set of semantics, sentics, and polarities associated with 100,000 natural language concepts. SenticNet assigns a polarity score in $[-1,1]$ for the concepts from negative to positive. The first step is to split a company’s $\ell$th news article, $d_\ell$, into sentences according to punctuations such as periods and question marks on day $t$ for $\ell=1,\ldots,N_t$. The second step is to select relevant sentences containing items such as stock symbols, company names, and product names. The selected sentences for $d_\ell$ are denoted by $s_{\ell j}$, $j=1,\ldots,n_\ell$, where $j$ denotes the $j$th sentence in $d_\ell$. The third step is to give a polarity score for each selected sentence using a weighted sum of the polarity scores of the $k$th phrases or words, denoted by $t_{\ell jk}$, in $s_{\ell j}$. When a negation is found in $s_{\ell j}$, the polarity of the whole sentence is inverted. In \cite{Metal2014}, the polarity of the sentence $s_{\ell j}$ is calculated as
	\begin{equation*}
		pol(s_{\ell j})=\sum_{k=1}^{|s_{\ell j}|}\frac{score(t_{\ell jk})\times w_{\ell jk}}{|s_{\ell j}|},
		\label{s_ij}
	\end{equation*}
	where $score(t_{\ell jk})$ is the polarity score of word $t_{\ell jk}$ assigned by SenticNet, $|s_{\ell j}|$ denotes the number of words in the sentence $s_{\ell j}$, and $w_{\ell jk}$ is defined by
	\begin{equation*}
		w_{\ell jk}=
		\left\{
		\begin{array}{rl}
			1.5, &{\rm if~} t_{\ell jk} \in\{{\rm adverbs,~verbs,~adjectives}\},\\
			1,   &{\rm otherwise.}
		\end{array}
		\right.
		\label{w_ijk}
	\end{equation*}
	The fourth step is to compute the polarity score of $d_\ell$ by $pol(d_\ell)=\sum_{j=1}^{n_\ell}pol(s_{\ell j}).$ Accordingly, let the score of soft information on day $t$, denoted by $D_{2,t}$, be the average of $pol(d_\ell)$, $\ell=1,\ldots,N_t$, that is
	\begin{equation}
		D_{2,t}=\frac{1}{N_t}\sum_{\ell=1}^{N_t}pol(d_\ell).
		\label{D0_t}
	\end{equation}
	To standardize the values of $D_{1,t}$ and $D_{2,t}$, we normalize them to $\tilde{D}_{1,t}$ and $\tilde{D}_{2,t}$, respectively, by the corresponding sample mean and sample standard deviation of the training data, and aggregate them by
	\begin{equation}
		D_t=\tilde{D}_{1,t}+\tilde{D}_{2,t}
		\label{D_t}
	\end{equation}
	to quantify the soft information. Notably, we assign equal weights to the normalized macroeconomic soft information $\tilde{D}_{1,t}$ and individual soft information $\tilde{D}_{2,t}$ when defining the score of soft information $D_t$ in equation (\ref{D_t}). While different weights could be applied by introducing a weight parameter or even a dynamic weight parameter in equation (\ref{D_t}), this would increase the computational complexity of the estimation procedure.

	\section{The Proposed SH-MBS-GARCH Model}
	\label{model}
	The proposed SH-MBS-GARCH model is a class of models that integrates hard and soft information to capture the joint dynamics of multiple financial time series, incorporating hysteretic effects within the MBS framework and addressing conditional heteroscedasticity through GARCH components. This section outlines its four specific models, each utilizing soft information from the EPU index and daily financial news in distinct ways to define the regime. Specifically, the regime indicator $R_t$ in equation (\ref{HAR2}) is formulated differently based on the incorporation of soft information.
	
	

	Assume that the multivariate time series ${\boldsymbol y}_t=({y}_{1,t},\ldots,{y}_{m,t})^\top$ follows the model
	\begin{eqnarray}
		&&{\boldsymbol y}_t = \boldsymbol{\mu}_t+\boldsymbol{\kappa}_t+\boldsymbol{\xi}^{(j)}_t+\boldsymbol{\varepsilon}_t,
		\label{H-MBS-GARCH}	
	\end{eqnarray}
	where the regime indicator $R_t = j$ for $j \in \{0, 1\}$. For $t = 1, 2, \ldots$, the components $(\boldsymbol{\mu}_t, \boldsymbol{\kappa}_t)$ are defined as in equation~(\ref{MBS}), and the error terms $\boldsymbol{\varepsilon}_t$ are i.i.d. with distribution $N_m(\mathbf{0}, \Sigma_\varepsilon)$. As in equation (\ref{xi}), we assume that the regression term $\boldsymbol{\xi}^{(j)}_t$ in equation (\ref{H-MBS-GARCH}) takes the form  $\boldsymbol{\xi}^{(j)}_t=\{(\boldsymbol{\beta}^{(j)}_i)^\top\boldsymbol{z}_{i,t}\}_{i=1}^m$, where  $\boldsymbol{\beta}^{(j)}_i$ is the vector associated with regression coefficients of the specified $k_i$-dimensional covariate $\boldsymbol{z}_{i,t}$ for the $i$th time series at time $t$.
	The regime indicator $R_t$ in equation (\ref{H-MBS-GARCH}) is defined as
	\begin{eqnarray}
		&&R_t = \left\{
		\begin{array}{ll}
			1  ,\qquad& \text{if } \sum_{i=1}^m\mathbbm{1}_{\{R_{i,t}=1\}}\ge k^\star m,\\
			0  ,& \text{if } \sum_{i=1}^m\mathbbm{1}_{\{R_{i,t}=0\}}\ge k^\star m,\\
			R_{t-1}  ,& \text{otherwise},
		\end{array}\right.
		\label{SHAR-GARCH1_4}
	\end{eqnarray}
	with $\mathbbm{1}_{A}$ being the indicator function of the event $A$, and $k^*$ in $[0.5,1]$ is a prespecified constant. A larger $k^*$ indicates a larger hysteresis zone. Equation (\ref{SHAR-GARCH1_4}) defines $R_{i,t}$ to flexibly incorporate both hard and soft information, with four specific approaches outlined below. To capture joint effects across $m$ assets, we model them simultaneously. However, with $2^m$ possible combinations of $(R_{1,t}, \ldots, R_{m,t})$, it becomes impractical to specify unique structures for each combination as $m$ grows. To address this, we propose a regime indicator at time $t$ based on the proportion of $R_{i,t}$ values equal to 1 (or 0) in equation (\ref{SHAR-GARCH1_4}). Specifically, if more than two-thirds of the $R_{i,t}$ values are 1 (or 0), we set $R_t = 1$ (or 0); otherwise, $R_t = R_{t-1}$.
	
	Since model equation (\ref{H-MBS-GARCH}) incorporates both soft information and GARCH effects within a hysteretic framework featuring an MBS structure in each regime, we refer to it as the SH-MBS-GARCH model. The first approach to defining $R_{i,t}$ relies exclusively on hard information.
	Therefore, similar to equation (\ref{HAR2}), let
	\begin{eqnarray}
		&&R_{i,t} = \left\{
		\begin{array}{ll}
			1  ,& \text{if } \tilde{r}_{i,t-1}< \tau^h_{i,L},\\
			0  ,& \text{if } \tilde{r}_{i,t-1}> \tau^h_{i,U},\\
			R_{i,t-1}  ,& \text{otherwise},
		\end{array}\right.
		\label{SHAR-GARCH1_4d}
	\end{eqnarray}
	where $\tilde{r}_{i,t}$ is an ${\cal F}_t$-measurable variable that depends on hard information, with $\tau^h{i,L}$ and $\tau^h_{i,U}$ representing the threshold parameters for the hard information of the $i$th asset, satisfying $\tau^h_{i,L} \leq \tau^h_{i,U}$, for $i = 1, \ldots, m$. This specific formulation is referred to as Type I of the SH-MBS-GARCH model.
	
	To incorporate a stronger influence of soft information in defining $R_{i,t}$, the Type II SH-MBS-GARCH model defines $R_{i,t}$ as follows:
	\begin{equation}
		R_{i,t} = \begin{cases} 
			1, & \text{if } \tilde{r}_{i,t-1} < \tau^h_{i,L} \text{ and } D_{i,t-1} < \tau^s_{i,U}, \\
			0, & \text{if } \tilde{r}_{i,t-1} > \tau^h_{i,U} \text{ and } D_{i,t-1} > \tau^s_{i,L}, \\
			R_{i,t-1}, & \text{otherwise},
		\end{cases}
		\label{SHAR-GARCH1_4a}
	\end{equation}
	where $\tilde{r}_{i,t}$, $\tau^h_{i,L}$, and $\tau^h_{i,U}$ are defined the same as in equation (\ref{SHAR-GARCH1_4d}), $D_{i,t}$ is defined in equation \eqref{D_t}, and $\tau^s_{i,L}$ and $\tau^s_{i,U}$ are threshold parameters for the soft information of the $i$th asset satisfying $\tau^s_{i,L} \leq \tau^s_{i,U}$, for $i = 1, \ldots, m$. Compared to equation \eqref{SHAR-GARCH1_4d}, the definition of $R_{i,t}$ in equation \eqref{SHAR-GARCH1_4a} assigns regions with contradictory information from hard and soft sources at time $t-1$ to the hysteresis zone. For instance, the region $\{(\tilde{r}_{i,t-1}, D_{i,t-1}) \mid \tilde{r}_{i,t-1} \leq \tau^h_{i,L} \text{ and } D_{i,t-1} > \tau^s_{i,U}\}$ reflects conflicting signals, where the hard information $\tilde{r}_{i,t-1}$ indicates low returns, while the soft information $D_{i,t-1}$ suggests a positive market trend.
	
	Using the same notation as in equations \eqref{SHAR-GARCH1_4d} and \eqref{SHAR-GARCH1_4a}, the Type III SH-MBS-GARCH model defines $R_{i,t}$ as follows:
	\begin{eqnarray}
		&&R_{i,t} = \left\{
		\begin{array}{ll}
			1  ,& \text{if } (\tilde{r}_{i,t-1},D_{i,t-1}) \in A_{i,t-1},\\
			0  ,& \text{if } (\tilde{r}_{i,t-1},D_{i,t-1}) \in B_{i,t-1}, \\
			R_{i,t-1}  ,& \text{otherwise},
		\end{array}\right.
		\label{SHAR-GARCH2_4a}
	\end{eqnarray}
	where 
	$$A_{i,t-1}=\Big\{(\tilde{r}_{i,t-1},D_{i,t-1})\mid(\tilde{r}_{i,t-1}\leq \tau^h_{i,L}) \cap (D_{i,t-1} \leq \tau^s_{i,U}) {\rm~or~} (\tilde{r}_{i,t-1}\leq \tau^h_{i,U}) \cap (D_{i,t-1} \leq \tau^s_{i,L}) \Big\},$$
	$$B_{i,t-1}=\Big\{(\tilde{r}_{i,t-1},D_{i,t-1})\mid(\tilde{r}_{i,t-1}>\tau^h_{i,U}) \cap (D_{i,t-1} > \tau^s_{i,L}) {\rm~or~} (\tilde{r}_{i,t-1}>\tau^h_{i,L}) \cap (D_{i,t-1} > \tau^s_{i,U}) \Big\}.$$ 
	Compared to equation \eqref{SHAR-GARCH1_4a}, the definition of $R_{i,t}$ in equation \eqref{SHAR-GARCH2_4a} expands two non-hysteresis zones, indicating that soft information plays a more significant role in the Type III SH-MBS-GARCH model than in the Type II SH-MBS-GARCH model.
	
\begin{figure}[!h]
		\begin{center}
			\includegraphics[scale=0.55]{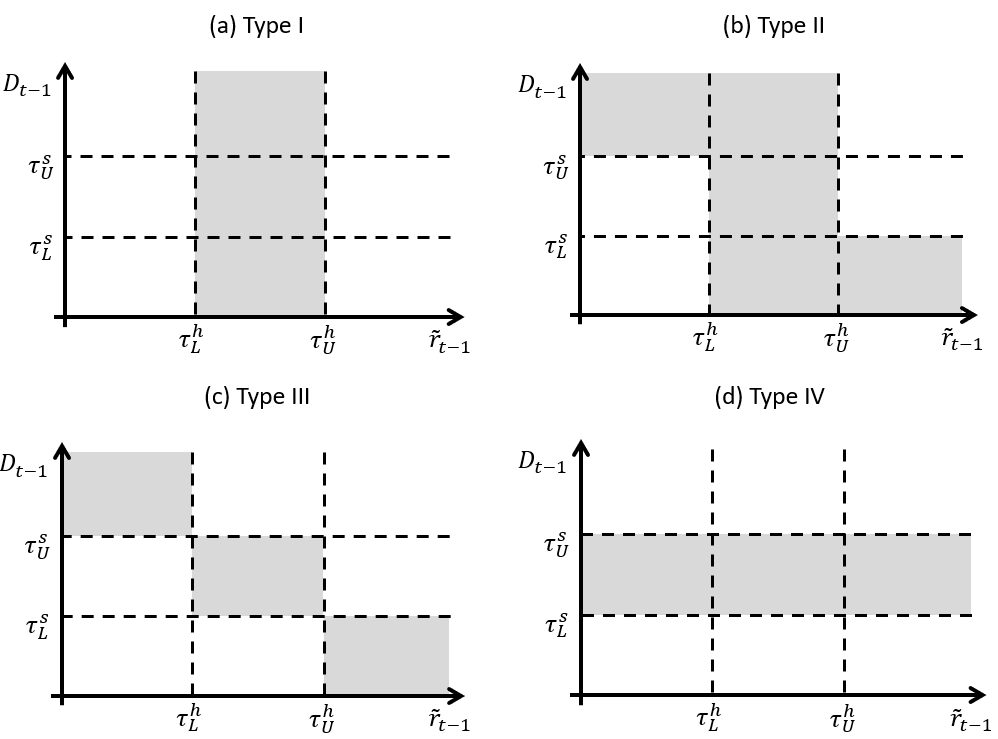}
			\caption{Hysteretic zones (gray regions) for the four types of the SH-MBS-GARCH model.}
			\label{fig1}
		\end{center}
	\end{figure}

	Finally, the Type IV SH-MBS-GARCH model defines $R_{i,t}$ as
	\begin{eqnarray}
		&&R_{i,t} = \left\{
		\begin{array}{ll}
			1  ,& \text{if } D_{i,t-1}\leq \tau^s_{i,L},\\
			0  ,& \text{if } D_{i,t-1} > \tau^s_{i,U}\\
			R_{i,t-1}  ,& \text{otherwise},
		\end{array}\right.
		\label{SHAR-GARCH3_4a}
	\end{eqnarray}
	where $R_{i,t}$ is determined solely by soft information. This formulation indicates that the Type IV SH-MBS-GARCH model relies more heavily on soft information to define the hysteresis zone compared to the previous three models. Figure \ref{fig1} illustrates the hysteretic zones for all four types of the SH-MBS-GARCH model.
	
	Let $\tau_{i,j}^k$ represent the $q_j^k$-th quantile of the conditional distribution of $\tilde{r}_{i,t}$ given $\mathcal{F}_{t-1}$ if $k = h$, or of the distribution of $D_{i,t}$ if $k = s$, for the $i$th asset, where $i = 1, \ldots, m$, $t = 1, \ldots, T$, and $j \in \{L, U\}$. Since the quantile $q_j^k$ is independent of $i$, this approach significantly reduces the number of parameters to be estimated. Once the conditional distribution of $\tilde{r}_{i,t}$ given $\mathcal{F}_{t-1}$ and the distribution of $D_{i,t}$ are known, $\tau_{i,j}^k$ can be expressed in terms of $q_j^k$. The conditional distribution of $\tilde{r}_{i,t}$, representing the de-GARCH return, can be directly derived from the specified GARCH-type model. We assume that $\{D_{i,t}\}_{t = 1, 2, \ldots}$ are i.i.d. random variables, and their distribution is estimated nonparametrically using the information in $\mathcal{F}_t$.

	\section{Parameter Estimation via Bayesian Inference}
	\label{sec:Parameter}
	
	In the SH-MBS-GARCH model, the regime-switching indicator $R_t$ results in path-dependent likelihood calculations, where the likelihood at time $t$ depends on the entire regime history up to that point. With multiple regimes, the number of possible regime paths grows exponentially with data length, significantly increasing the computational burden of likelihood evaluation. To address this, we propose a Bayesian estimation approach using MCMC sampling. We enhance efficiency by adopting the adaptive MCMC method of \cite{CS2006}, which efficiently explores the high-dimensional, path-dependent posterior distribution. This approach mitigates strong posterior correlations and improves convergence, particularly in complex regime-switching models \citep{CT2016,CC2024}. 
	
	In high-dimensional multivariate time series models, the parameter space grows rapidly with the number of variables and lags, risking overparameterization and high computational complexity \citep{qiu2020multivariate}. Variable selection is crucial to ensure model parsimony, enhance interpretability, and improve computational efficiency by identifying key components driving system dynamics. To achieve this, we apply spike-and-slab priors \citep{GM1993,GM1997} to a regression component, facilitating automatic structure learning and promoting posterior sparsity. This approach shrinks irrelevant coefficients toward zero, enabling the model to focus on statistically and substantively significant predictors.

	Let $\boldsymbol{\theta}$ denote the full parameter vector for the proposed model, as defined in equation (\ref{H-MBS-GARCH}). We partition $\boldsymbol{\theta}$ into four components: $\boldsymbol{\theta} = (\boldsymbol{\theta}_1, \boldsymbol{\theta}_2, \boldsymbol{\theta}_3, \boldsymbol{\theta}_4)$, where $\boldsymbol{\theta}_1 = (q_L^h, q_U^h, q_L^s, q_U^s)$ represents the quantile bounds for the regime indicators, $\boldsymbol{\theta}_2 = (\boldsymbol{\mu}_t, \boldsymbol{\kappa}_t)$ includes the nonstationary components, $\boldsymbol{\theta}_3 = (\Sigma_u, \Sigma_v, \Sigma_w)$ comprises the covariance matrices associated with $\boldsymbol{\mu}_t$ and $\boldsymbol{\kappa}_t$, and $\boldsymbol{\theta}_4 = (\boldsymbol{\beta}, \Sigma_\varepsilon)$ contains the regression coefficients in $\boldsymbol{\xi}^{(j)}_t$ for $j=0,1$ and the covariance matrix of $\boldsymbol{\varepsilon}_t$. The proposed MCMC procedure, outlined in Algorithm \ref{alg1}, integrates adaptive MCMC techniques to enhance convergence and employs spike-and-slab priors for variable selection to identify relevant predictors. Brief overviews of adaptive MCMC and spike-and-slab regression are provided in \ref{secA.1} and \ref{secA.2}, respectively. 
	
	\begin{algorithm}[t!]
		\caption{Bayesian Estimation for SH-MBS-GARCH Model via MCMC}\label{alg:SH-MBS-GARCH}
		\begin{enumerate}
			\bigskip
			
			\item Initialize the parameter vectors $\boldsymbol{\theta}_1$, $\boldsymbol{\theta}_2$, $\boldsymbol{\theta}_3$, and $\boldsymbol{\theta}_4$ with their respective starting values: $\boldsymbol{\theta}_1^{[0]}$, $\boldsymbol{\theta}_2^{[0]}$, $\boldsymbol{\theta}_3^{[0]}$, and $\boldsymbol{\theta}_4^{[0]}$.
			
			\item For iteration $\ell = 1, \dots, 1500$, perform the following steps:
			\begin{enumerate}
				\item Sample $\boldsymbol{\theta}_1$ from the conditional posterior $p(\boldsymbol{\theta}_1 \mid \boldsymbol{y}, \boldsymbol{\theta}_2, \boldsymbol{\theta}_3, \boldsymbol{\theta}_4)$ using the adaptive MCMC method, as specified in equation (\ref{hysterestic}).
				
				\item Simulate the latent state $\boldsymbol{\theta}_2$ from $p(\boldsymbol{\theta}_2 \mid \boldsymbol{y}, \boldsymbol{\theta}_1, \boldsymbol{\theta}_3, \boldsymbol{\theta}_4)$ using the simulation smoother method of \cite{DK2002}. See \ref{secA.4} for details.
				
				\item Sample $\boldsymbol{\theta}_3$ from the conditional posterior $p(\boldsymbol{\theta}_3 \mid \boldsymbol{y}, \boldsymbol{\theta}_1, \boldsymbol{\theta}_2, \boldsymbol{\theta}_4)$ using the Gibbs sampling algorithm, based on equation (\ref{sigma_u}) in Section \ref{sec4.3}.
				
				\item Sample $\boldsymbol{\theta}_4$ from the conditional posterior $p(\boldsymbol{\theta}_4 \mid \boldsymbol{y}, \boldsymbol{\theta}_1, \boldsymbol{\theta}_2, \boldsymbol{\theta}_3)$ using the spike-and-slab regression method, as described in equations (\ref{beta_post2})--(\ref{gamma_post}) in Section \ref{sec4.3}.
			\end{enumerate}
			
			\item Output $\boldsymbol{\theta}_1^{[1500]}$, $\boldsymbol{\theta}_2^{[1500]}$, $\boldsymbol{\theta}_3^{[1500]}$, and $\boldsymbol{\theta}_4^{[1500]}$.
		\end{enumerate}
		\label{alg1}
	\end{algorithm}
	
	\subsection{Prior distributions}
	\label{sec:Prior}
	
	We specify the prior distributions for the parameter vectors $\boldsymbol{\theta}_1$, $\boldsymbol{\theta}_2$, $\boldsymbol{\theta}_3$, and $\boldsymbol{\theta}_4$ used in Algorithm \ref{alg:SH-MBS-GARCH}. The prior for $\boldsymbol{\theta}_1 = (q_{L}^h, q_{U}^h, q_{L}^s, q_{U}^s)$ is defined as:
	\begin{align}
		q_{L}^h &\sim \text{Unif}(a_1, b_1), \qquad 
		q_{U}^h \mid q_{L}^h \sim \text{Unif}(a_2, b_2), \nonumber \\
		q_{L}^s &\sim \text{Unif}(c_1, d_1), \qquad 
		q_{U}^s \mid q_{L}^s \sim \text{Unif}(c_2, d_2),
		\label{theta1 prior}
	\end{align}
	where $a_1$ and $b_1$ are the $\varphi_h$-th and $\varphi_{1-2h}$-th percentiles of the observations, respectively. The bounds for $q_{U}^h$ are set as $a_2 = q_{L}^h + c^*$ and $b_2 = \varphi_{1-h}$, where $c^*$ is a pre-specified constant ensuring $q_{L}^h + c^* < q_{U}^h$ and that at least $100h\%$ of observations lie within $(q_{L}^h, q_{U}^h)$. Similarly, $c_1$ and $d_1$ are the $\phi_h$-th and $\phi_{1-2h}$-th percentiles of another observation set, with bounds for $q_{U}^s$ defined as $c_2 = q_{L}^s + c^*$ and $d_2 = \phi_{1-h}$, following the same construction as for the hard-information bounds. We employ data-dependent priors for $(q_L, q_U)$, with hyperparameters derived from sample-based summary statistics. This approach allows the priors to adapt flexibly to the observed data structure, enhancing modeling versatility in high-dimensional or complex scenarios. The use of data-dependent priors is well-supported in the literature, notably by \citet{Richardson1997} and  \citet{W2000}, who highlight their practicality and effectiveness in complex Bayesian models. 
	
	For the prior distribution of $\boldsymbol{\theta}_2 = (\boldsymbol{\mu}, \boldsymbol{\kappa})$, we adopt the dynamic Gaussian state-space modeling framework of \citet{DK2002}. This approach is advantageous for enabling efficient and asymptotically valid inference, even with diffuse priors \citep{Scott2014,Qetal2018}. Accordingly, we implement this framework to initialize and update $\boldsymbol{\theta}_2$ within Algorithm \ref{alg:SH-MBS-GARCH}. Specifically, the initial prior for $\boldsymbol{\theta}_2$ in Step 1 of Algorithm \ref{alg:SH-MBS-GARCH} is specified as:
	\[
	\boldsymbol{\theta}_2 \sim \mathcal{N}(\boldsymbol{a}_1, \boldsymbol{P}_1),
	\]
	where $\boldsymbol{a}_1$ and $\boldsymbol{P}_1$ are known and can be chosen to reflect either informative or diffuse beliefs about the initial latent state.
	
	For the variance components in $\boldsymbol{\theta}_3 = (\Sigma_u, \Sigma_v, \Sigma_w)$, we specify the following prior:
	\begin{equation}
		\Sigma_a \sim \text{IW}(w_a, W_a), \qquad a \in \{u, v, w\},
		\label{theta3 prior}
	\end{equation}
	where $\text{IW}(w_a, W_a)$ denotes the inverse Wishart distribution with $w_a$ degrees of freedom and scale matrix $W_a \in \mathbb{R}^{m \times m}$. This choice serves as a conjugate prior for the covariance matrices of the multivariate normal innovations in the trend and seasonal components. To simplify computation and reflect the assumption that state components (e.g., trend and seasonality) evolve independently across series, we specify diagonal scale matrices $W_a$. Consequently, the inverse Wishart prior decomposes into independent inverse gamma distributions for each diagonal element. This structure aligns with the prior framework commonly used in Bayesian Structural Time Series (BSTS) models \citep{Scott2014}, facilitating independent modeling of component-specific variances while maintaining conjugacy for efficient posterior inference.
	
	For $\boldsymbol{\theta}_4 = (\boldsymbol{\beta}, \Sigma_\varepsilon)$, we adopt a spike-and-slab prior by introducing a binary inclusion vector $\boldsymbol{\gamma} = (\gamma_{1,1}, \dots, \gamma_{1,k_1}, \dots, \gamma_{m,1}, \dots, \gamma_{m,k_m})^\top$, where each $\gamma_{ij} \sim \operatorname{Ber}(\pi_{ij})$ with $0 \leq \pi_{ij} \leq 1$. If $\gamma_{ij} = 1$, the coefficient $\boldsymbol{\beta}_i^{(j)} \in \boldsymbol{\theta}_4$ is deemed relevant; otherwise, it is considered irrelevant. Following \citet{G2003}, we assume prior independence between $\boldsymbol{\beta}$ and $\Sigma_\varepsilon$, specifying the joint prior as:
	\begin{equation}
		p(\boldsymbol{\beta}, \Sigma_\varepsilon, \boldsymbol{\gamma}) = 
		p(\boldsymbol{\beta} \mid \boldsymbol{\gamma}) \cdot 
		p(\Sigma_\varepsilon \mid \boldsymbol{\gamma}) \cdot 
		p(\boldsymbol{\gamma}),
		\label{prior3}
	\end{equation}
	with component priors defined as:
	\begin{equation}
		p(\boldsymbol{\gamma}) = \prod_{i=1}^{m} \prod_{j=1}^{k_i} 
		\pi_{ij}^{\gamma_{ij}} (1 - \pi_{ij})^{1 - \gamma_{ij}}, \qquad
		\Sigma_\varepsilon \mid \boldsymbol{\gamma} \sim \mathrm{IW}(\phi, \nu), \qquad
		\boldsymbol{\beta} \mid \boldsymbol{\gamma} \sim \mathcal{N}(c_{\gamma}, D_{\gamma}^{-1}),
		\label{prior beta}
	\end{equation}
	where $D_\gamma = \psi Z_\gamma^\top Z_\gamma / n$ is the prior information matrix for the full model, with $\psi$ representing the weight (in terms of observation count) assigned to the prior mean vector. The vector $c_{\gamma}$ encodes prior beliefs about the active coefficients $\beta_{\gamma}$, typically set to $c_{\gamma} = 0$ in practice.

	In equations (\ref{prior3}) and (\ref{prior beta}), the marginal prior distribution $p(\boldsymbol{\gamma})$ is termed the ``spike'' prior, as it assigns significant probability mass to zero, promoting sparsity in model selection. In the absence of specific prior knowledge about individual predictors, we assume a common inclusion probability for all predictors within each asset, defined as $\pi_{ij} = p_i / k_i$, where $p_i$ represents the expected number of nonzero predictors for the $i$-th asset, and $k_i$ is the total number of candidate predictors. The conditional priors $p(\boldsymbol{\beta} \mid \boldsymbol{\gamma})$ and $p(\Sigma_\varepsilon \mid \boldsymbol{\gamma})$ constitute the ``slab'' component, with parameters that can be set to be weakly informative (i.e., nearly flat) given the inclusion configuration $\boldsymbol{\gamma}$.

	\subsection{Posterior distributions}
	\label{sec4.3}
	By Bayes’ theorem, the conditional posterior distribution for each parameter block is proportional to the product of the likelihood function,
	denoted by $L(\boldsymbol{y} \mid \boldsymbol{\theta_1}, \boldsymbol{\theta_2}, \boldsymbol{\theta_3},\boldsymbol{\theta_4})$, and the associated prior distributions. The conditional posterior distributions for each parameter group are derived and summarized below. The conditional posterior distribution for \(\boldsymbol{\theta}_1\) is given by
	\begin{equation}
		p(\boldsymbol{\theta}_1 \mid \boldsymbol{y}_t, \boldsymbol{\theta}_2, \boldsymbol{\theta}_3, \boldsymbol{\theta}_4) 
		\propto 
		L(\boldsymbol{y}_t \mid \boldsymbol{\theta}_1, \boldsymbol{\theta}_2, \boldsymbol{\theta}_3,\boldsymbol{\theta}_4) \times p(\boldsymbol{\theta}_1),
		\label{hysterestic}
	\end{equation}
	where the prior \( p(\boldsymbol{\theta}_1) = p(q^{k}_{i,L}) \times p(q^{k}_{U} \mid q^{k}_{L}) \)  is specified in equation (\ref{theta1 prior}), and
	\begin{align}
		\log L(\boldsymbol{y} \mid \boldsymbol{\theta}_1, \boldsymbol{\theta}_2, \boldsymbol{\theta}_3, \boldsymbol{\theta}_4)
		&= c
		-\frac{1}{2} \sum_{t=2}^{n} \sum_{j=0}^{1} \mathbbm{1}_{\{R_{t}=j\}}
		\Big\{
		\log\lvert\Sigma_{\varepsilon}\rvert
		+ \bigl(\mathbf{e}_{t}^{(j)}\bigr)^{\top}
		\bigl(\Sigma_{\varepsilon}\bigr)^{-1}
		\mathbf{e}_{t}^{(j)}
		\Big\},
		\label{L}
	\end{align}
	with $c$ being a constant and $\mathbf e_{t}^{(j)}
	= \boldsymbol{y}_{t}-\boldsymbol{\mu}_{t}-\boldsymbol{\kappa}_{t}-\boldsymbol{\beta}_1^{(j)}\boldsymbol{z}_{t-1}-\boldsymbol{\beta}_2^{(j)}\boldsymbol{z}_{t-2}-\boldsymbol{\beta}_3^{(j)}\boldsymbol{z}_{t-3}$.
	The structure of the conditional posterior distribution adopted in (\ref{hysterestic}) is consistent with established practices in the Bayesian regime-switching literature, as demonstrated in \citep{CT2016,CTSS2019,CC2024}.
	
	The conditional posterior distribution for $\boldsymbol{\theta}_3 = (\Sigma_u, \Sigma_v, \Sigma_w)$ is derived using the conjugate relationship between the multivariate normal likelihood and the inverse Wishart prior specified in equation (\ref{theta3 prior}). Assuming the residuals of each time series component follow a multivariate normal distribution, the posterior distribution for each covariance matrix is:
	\begin{equation}
		\Sigma_a \mid \boldsymbol{y}, \boldsymbol{\theta}_1, \boldsymbol{\theta}_2, \boldsymbol{\theta}_4 \sim \operatorname{IW}(w_a + n, W_a + A A^\top), \qquad a \in \{u, v, w\},
		\label{sigma_u}
	\end{equation}
	where $A = [\tilde{A}_1, \ldots, \tilde{A}_n]$ is an $m \times n$ matrix collecting the residuals for each time series component, derived from draws of the latent states and representing the unexplained variation in the corresponding component.

	Let $\boldsymbol{Y}= \boldsymbol{y}-\boldsymbol{\mu}-\boldsymbol{\kappa} = \boldsymbol{Z}\boldsymbol{\beta} + 
	\boldsymbol{\varepsilon}$
	be the target series after removing trend and seasonal components in
	(\ref{H-MBS-GARCH}). Under the conditional independence between $\boldsymbol{\beta}$ and $\Sigma_\varepsilon$ given $\boldsymbol{\gamma}$ in equation (\ref{prior3}), the joint density of $(\boldsymbol{Y}, \boldsymbol{\beta}, \Sigma_\varepsilon, \boldsymbol{\gamma})$ is given by
	\begin{equation}
		p(\boldsymbol{Y}, \boldsymbol{\beta}, \Sigma_\varepsilon, \boldsymbol{\gamma})
		= L(\boldsymbol{Y} \mid \boldsymbol{\beta}, \Sigma_\varepsilon, \boldsymbol{\gamma})
		\cdot p(\boldsymbol{\beta} \mid \boldsymbol{\gamma})
		\cdot p(\Sigma_\varepsilon \mid \boldsymbol{\gamma})
		\cdot p(\boldsymbol{\gamma}),
		\label{fullL}
	\end{equation}
	where the likelihood function $L(\boldsymbol{Y} \mid \boldsymbol{\beta}, \Sigma_\varepsilon, \boldsymbol{\gamma})$ assumes Gaussian observation errors:
	\begin{equation}
		L(\boldsymbol{Y} \mid \boldsymbol{\beta}, \Sigma_\varepsilon, \boldsymbol{\gamma})
		\propto
		|\Sigma_\varepsilon|^{-n/2}
		\exp\left\{
		-\frac{1}{2}
		(\boldsymbol{Y} - Z_{\gamma}\boldsymbol{\beta}_{\gamma})^{\top}
		(\Sigma_\varepsilon^{-1} \otimes I_n)
		(\boldsymbol{Y} - Z_{\gamma}\boldsymbol{\beta}_{\gamma})
		\right\},
		\label{condition y}
	\end{equation}
	and $p(\boldsymbol{\beta} \mid \boldsymbol{\gamma})$, $p(\Sigma_\varepsilon \mid \boldsymbol{\gamma})$, and $p(\boldsymbol{\gamma})$ are as specified in equation (\ref{prior beta}). In equation (\ref{condition y}), $\boldsymbol{\beta}_\gamma$ denotes the subset of $\boldsymbol{\beta}$ where $\boldsymbol{\beta}_{ij} \neq 0$, $Z_\gamma$ represents the columns of $Z$ corresponding to $\gamma_{ij} = 1$, and $\otimes$ is the Kronecker product operator.
	
	The conditional posterior distribution of $\boldsymbol{\beta}$ given $(\boldsymbol{Y}, \Sigma_\varepsilon, \boldsymbol{\gamma})$ is:
	\begin{equation}
		p(\boldsymbol{\beta} \mid \boldsymbol{Y}, \Sigma_\varepsilon, \boldsymbol{\gamma}) \propto p(\boldsymbol{Y} \mid \boldsymbol{\beta}, \Sigma_\varepsilon, \boldsymbol{\gamma}) \cdot p(\boldsymbol{\beta} \mid \boldsymbol{\gamma}),
		\label{beta_post0}
	\end{equation}
	To derive this, we apply a transformation technique using the Cholesky decomposition of $\Sigma_\varepsilon$, such that $\Sigma_\varepsilon = U^\top U$. By multiplying both sides of $\boldsymbol{Y} = \boldsymbol{Z} \boldsymbol{\beta} + \boldsymbol{\varepsilon}$ by $\Psi = (U^{-1})^\top \otimes I_n$, we obtain:
	\begin{equation}
		\boldsymbol{Y}^\star = \boldsymbol{Z}^\star \boldsymbol{\beta} + \boldsymbol{\varepsilon}^\star,
		\label{Ystar}
	\end{equation}
	where $\boldsymbol{Y}^\star = \Psi \boldsymbol{Y}$, $\boldsymbol{Z}^\star = \Psi \boldsymbol{Z}$, and $\boldsymbol{\varepsilon}^\star = \Psi \boldsymbol{\varepsilon}$. Given $\boldsymbol{\varepsilon} \sim \mathcal{N}(0, \Sigma_\varepsilon)$, the variance of the transformed error term $\boldsymbol{\varepsilon}^\star$ is:
	\begin{align*}
		\operatorname{Var}(\boldsymbol{\varepsilon}^\star) 
		&= \operatorname{Var}(\Psi \boldsymbol{\varepsilon}) 
		= \Psi (\Sigma_\varepsilon \otimes I_n) \Psi^\top 
		= \bigl( (U^{-1})^\top \otimes I_n \bigr) (U^\top U \otimes I_n) \bigl( U^{-1} \otimes I_n \bigr)\\
		&= (U^{-1})^\top U^\top U U^{-1} \otimes I_n 
		= I_m \otimes I_n,
	\end{align*}
	where the result follows from the matrix identity $$(A \otimes B)(C \otimes D) = (AC) \otimes (BD).$$ Thus, the transformed error term is $\boldsymbol{\varepsilon}^\star \sim \mathcal{N}(0, I_m \otimes I_n)$.

	The transformation in equation (\ref{Ystar}) results in uncorrelated and homoskedastic error terms across cross-sectional units and time periods, significantly enhancing posterior inference and computational efficiency. Consequently, the conditional posterior in equation (\ref{beta_post0}) can be reformulated as:
	\begin{equation}
		p(\boldsymbol{\beta} \mid \boldsymbol{Y}, \Sigma_\varepsilon, \boldsymbol{\gamma}) \propto p(\boldsymbol{Y}^\star \mid \boldsymbol{\beta}, \Sigma_\varepsilon, \boldsymbol{\gamma}) \cdot p(\boldsymbol{\beta} \mid \boldsymbol{\gamma}),
		\label{beta}
	\end{equation}
	where 
	$
	p(\boldsymbol{Y}^\star \mid \boldsymbol{\beta}, \Sigma_\varepsilon, \boldsymbol{\gamma}) \propto \exp\left( -\frac{1}{2} (\boldsymbol{Y}^\star - \boldsymbol{Z}^\star_\gamma \boldsymbol{\beta}_\gamma)^\top (\boldsymbol{Y}^\star - \boldsymbol{Z}^\star_\gamma \boldsymbol{\beta}_\gamma) \right),
	$
	with $\boldsymbol{Y}^\star = \Psi \boldsymbol{Y}$ as defined in equation (\ref{Ystar}), and $p(\boldsymbol{\beta} \mid \boldsymbol{\gamma})$ specified in equation (\ref{prior beta}). By direct computation, the right-hand side of equation (\ref{beta}) becomes:
	\[
	p(\boldsymbol{Y}^\star \mid \boldsymbol{\beta}, \Sigma_\varepsilon, \boldsymbol{\gamma}) \cdot p(\boldsymbol{\beta} \mid \boldsymbol{\gamma}) \propto \exp\left( -\frac{1}{2} (\boldsymbol{\beta}_\gamma - \tilde{\boldsymbol{\beta}}_\gamma)^\top (\boldsymbol{Z}_\gamma^{\star\top} \boldsymbol{Z}^\star_\gamma + D_\gamma) (\boldsymbol{\beta}_\gamma - \tilde{\boldsymbol{\beta}}_\gamma) \right),
	\]
	where 
	\[
	\tilde{\boldsymbol{\beta}}_\gamma = (\boldsymbol{Z}_\gamma^{\star\top} \boldsymbol{Z}^\star_\gamma + D_\gamma)^{-1} (\boldsymbol{Z}_\gamma^{\star\top} \boldsymbol{Y}^\star + D_\gamma c_\gamma).
	\]
	Thus, the conditional posterior distribution of $\boldsymbol{\beta}_\gamma$ is given by
	\begin{equation}
		\boldsymbol{\beta}_\gamma \mid \boldsymbol{Y}, \Sigma_\varepsilon, \boldsymbol{\gamma} \sim \mathcal{N} \left( \tilde{\boldsymbol{\beta}}_\gamma, (\boldsymbol{Z}_\gamma^{\star\top} \boldsymbol{Z}^\star_\gamma + D_\gamma)^{-1} \right).
		\label{beta_post2}
	\end{equation}

	Using equations (\ref{prior beta}), (\ref{fullL}), and (\ref{condition y}), and defining $E_\gamma = \boldsymbol{Y} - \boldsymbol{Z}_\gamma \boldsymbol{\beta}_\gamma$, the conditional posterior for $\Sigma_\varepsilon \mid \boldsymbol{Y}, \boldsymbol{\beta}, \boldsymbol{\gamma}$ is:
	\begin{align*}
		p(\Sigma_\varepsilon \mid \boldsymbol{Y}, \boldsymbol{\beta}, \boldsymbol{\gamma}) 
		&\propto L(\boldsymbol{Y} \mid \boldsymbol{\beta}, \Sigma_\varepsilon, \boldsymbol{\gamma}) \cdot p(\Sigma_\varepsilon \mid \boldsymbol{\gamma}) \\
		&\propto |\Sigma_\varepsilon|^{-(n + \phi + m + 1)/2} \exp\left( -\frac{1}{2} \operatorname{tr}\{ (E_\gamma^\top E_\gamma + \nu) \Sigma_\varepsilon^{-1} \} \right),
		\label{Sigma_post}
	\end{align*}
	indicating that the posterior distribution of $\Sigma_\varepsilon$ given $(\boldsymbol{Y}, \boldsymbol{\beta}, \boldsymbol{\gamma})$ follows an inverse Wishart distribution:
	\begin{equation}
		\Sigma_\varepsilon \mid \boldsymbol{Y}, \boldsymbol{\beta}, \boldsymbol{\gamma} \sim \operatorname{IW}(\phi + n, E_\gamma^\top E_\gamma + \nu),
	\end{equation}
	where $\phi + n$ represents the degrees of freedom, and $E_\gamma^\top E_\gamma + \nu$ is the scale matrix.

	The vector $\boldsymbol{\gamma}$ comprises indicator variables that determine the inclusion of model components (e.g., trends or seasonality). Its prior, typically non-conjugate, prevents closed-form computation of the marginal likelihood $p(\boldsymbol{\gamma} \mid \boldsymbol{Y}, \boldsymbol{\beta}, \Sigma_\varepsilon)$. Instead, we derive the posterior distribution of $\boldsymbol{\gamma}$ conditional on $\boldsymbol{Y}$ and $\Sigma_\varepsilon$ to facilitate inference.
	
	Using equations (\ref{fullL}) and the linear transformation in equation (\ref{Ystar}), the joint probability density $p(\boldsymbol{Y}, \Sigma_\varepsilon, \boldsymbol{\gamma})$ is:
	\begin{align*}
		p(\boldsymbol{Y}, \Sigma_\varepsilon, \boldsymbol{\gamma}) 
		&\propto \int_{-\infty}^{+\infty} 
		p(\boldsymbol{Y} \mid \boldsymbol{\beta}, \Sigma_\varepsilon, \boldsymbol{\gamma}) \, p(\boldsymbol{\beta} \mid \boldsymbol{\gamma}) \, p(\Sigma_\varepsilon \mid \boldsymbol{\gamma}) \, p(\boldsymbol{\gamma}) \,d\boldsymbol{\beta} \\
		&\propto 
		\frac{|\Sigma_\varepsilon|^{-(n + \phi + m + 1)/2} |D_\gamma|^{1/2} \, p(\boldsymbol{\gamma})}
		{\left| (\boldsymbol{Z}_\gamma^{\star\top}) \boldsymbol{Z}_\gamma^{\star} + D_\gamma \right|^{1/2}} 
		\exp\left( 
		-\frac{1}{2} \left\{ 
		\operatorname{tr}(V \Sigma_\varepsilon^{-1}) + \boldsymbol{Y}^{\star\top} \boldsymbol{Y}^{\star} 
		\right\} 
		\right) \\
		&\quad \times 
		\exp\left( 
		-\frac{1}{2} \left( 
		c_\gamma^{\top} D_\gamma c_\gamma 
		- P_\gamma^{\top} \left( (\boldsymbol{Z}_\gamma^{\star\top}) \boldsymbol{Z}_\gamma^{\star} + D_\gamma \right)^{-1} P_\gamma 
		\right) 
		\right),
	\end{align*}
	where $P_\gamma = \boldsymbol{Z}_\gamma^{\star\top} \boldsymbol{Y}^{\star} + D_\gamma c_\gamma$ and $\tilde{\boldsymbol{\beta}}_\gamma = \left( \boldsymbol{Z}_\gamma^{\star\top} \boldsymbol{Z}_\gamma^{\star} + D_\gamma \right)^{-1} P_\gamma$. Consequently, the conditional posterior distribution of $\boldsymbol{\gamma} \mid \boldsymbol{Y}, \Sigma_\varepsilon$ can be expressed as:
	\begin{align}
		&p(\boldsymbol{\gamma} \mid \boldsymbol{Y},\Sigma_\varepsilon) \nonumber\\
		&\propto
		\frac{|D_\gamma|^{1/2} p(\boldsymbol{\gamma})}{\left| (\boldsymbol{Z}_\gamma^{\star})^{\top} \boldsymbol{Z}_\gamma^{\star} + D_\gamma \right|^{1/2}} 
		\exp\left( -\frac{1}{2} \left( c_\gamma^{\top} D_\gamma c_\gamma - P_\gamma^{\top} ((\boldsymbol{Z}_\gamma^{\star})^{\top} \boldsymbol{Z}_\gamma^{\star} + D_\gamma)^{-1} P_\gamma \right) \right), \hspace*{-0.3cm}
		\label{gamma_post}
	\end{align}
	The sparsity induced by $\boldsymbol{\gamma}$ enhances the efficiency of evaluating this posterior. Each $\gamma_i$ is sampled sequentially in random order, conditioned on $\boldsymbol{\gamma}_{-i}, \boldsymbol{Y}$ and  $\boldsymbol{\theta}_1$, following equation (\ref{gamma_post}), using the stochastic search variable selection algorithm of \cite{GM1997}. The typically low-dimensional matrices in equation (\ref{gamma_post}) further simplify posterior inference.

	\section{Numerical Results}
	\label{sec:Numerical}
	\begin{table}[t!]
		\begin{center}
			\begin{tabular}{cr|cr|cr}
				Para. &value &Para. &value &Para. &value \\ \hline
				$\phi_{1,1}$&0.6 &$\phi_{2,1}$&0.8 &$\phi_{3,1}$&0.5  \\
				$\phi_{1,2}$&-0.8  &$\phi_{2,2}$&-0.6  &$\phi_{3,2}$&0.3  \\
				$\phi_{1,3}$&0.5  &$\phi_{2,3}$&0.5  &$\phi_{3,3}$&-0.7  \\
				$\sigma_1^2$&$0.5$  &$\sigma_2^2$&$0.5$   &$\sigma_3^2$&$0.5$  \\
				\hline
			\end{tabular}	
		\end{center}
		\caption{The parameter settings for equation (\ref{4.1}) in the simulation study}
		\label{table1}
	\end{table} 
	
	In this section, we present numerical results, encompassing a simulation study in Section \ref{sec:Simulation}, data description in Section \ref{sec:Data}, and an empirical study in Section \ref{sec:Empirical}. The associated code and data are available at \texttt{https://reurl.cc/GNxyKD}.
	
	\subsection{Simulation study}
	\label{sec:Simulation}
\begin{figure}[t!]
		\begin{center}
			\includegraphics[scale=1.1]{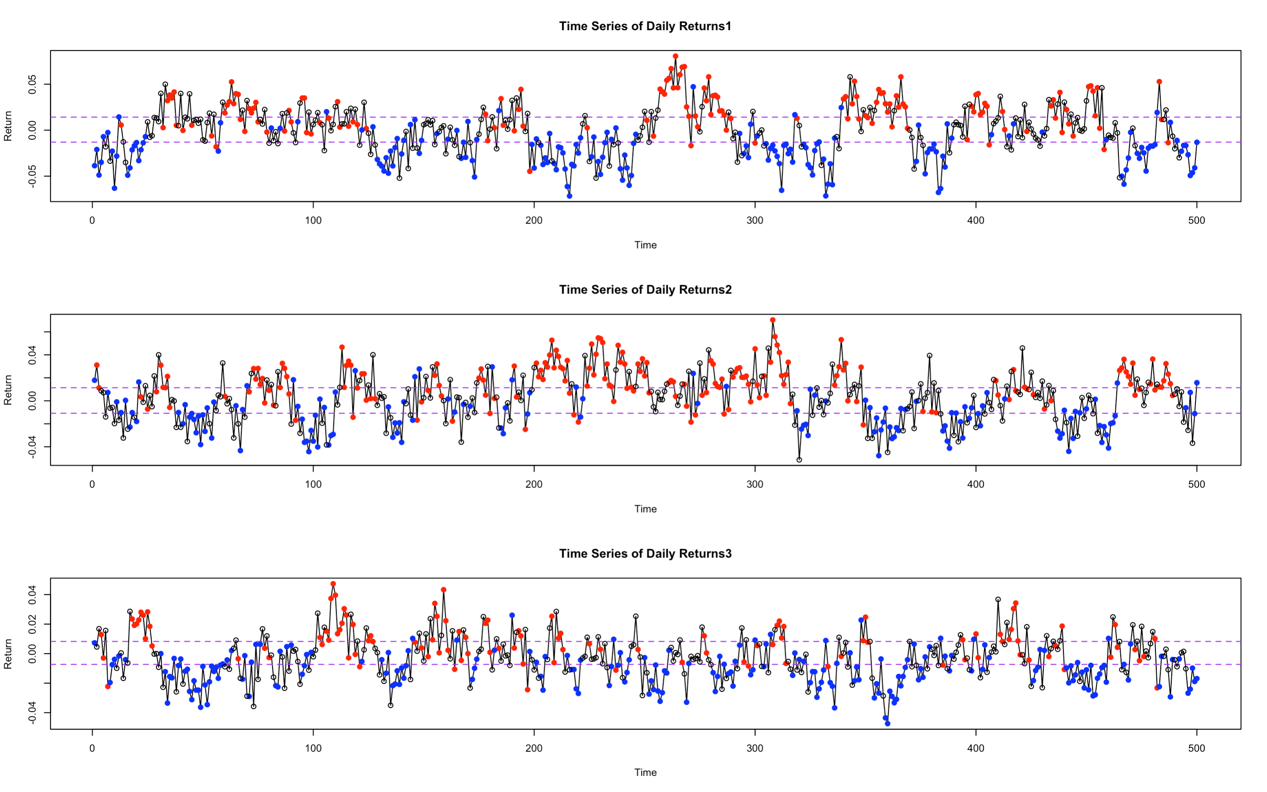}
			\caption{The three simulated AR(3) processes, with dashed lines indicating $\tau_{i,L}$ and $\tau_{i,U}$, and red and blue dots representing the two regimes defined by equation (\ref{HAR2}) with $d=1$ and $(\tau_{i,L}, \tau_{i,U})$ for $i=1,2,3$.}
			\label{fig02}
		\end{center}
	\end{figure}

	We conduct a simulation study for the SH-MBS-GARCH model, as defined in equation (\ref{H-MBS-GARCH}), with $m=3$ series, $k=3$ predictors and $k^\star =2/3$. For simplicity, we modify the model as follows. First, we omit the de-GARCH steps and generate the time series $\boldsymbol{x}_t = (x_{1,t}, x_{2,t}, x_{3,t})^\top$ using a stationary AR(3) process:
	\begin{equation}
		x_{i,t} = \sum_{p=1}^3 \phi_{i,p} x_{i,t-p} + \varepsilon_{i,t}, \qquad t=1,\ldots,n,
		\label{4.1}
	\end{equation}
	where $\{\phi_{i,p}, p=1,2,3\}$ satisfy the stationarity condition, and $\varepsilon_{i,t}$ for $t=1,\ldots,500$ are i.i.d. $\mathcal{N}(0, \sigma_i^2)$ innovations for $i=1,2,3$. The parameter settings are provided in Table \ref{table1}. Second, we focus solely on hard information in equation (\ref{SHAR-GARCH1_4a}), simplifying to the form in equation (\ref{HAR2}) with $d=1$, to define the regime component for each univariate time series $i$. The regime parameters $\tau_{i,L}$ and $\tau_{i,U}$ are set as the 30th and 70th percentiles of $\{x_{i,t}, t=1,\ldots,500\}$ for each $i \in \{1,2,3\}$. That is, $q_L=0.3$ and $q_U=0.7$. Figure \ref{fig02} displays the generated $x_{i,t}$ processes for $i=1,2,3$, with dashed lines indicating $\tau_{i,L}$ and $\tau_{i,U}$, and red and blue dots marking the two states defined by equation (\ref{HAR2}) with $d=1$ and $(\tau_{i,L}, \tau_{i,U})$. The regime $R_t$ in equation (\ref{H-MBS-GARCH}) is thus determined by equation (\ref{SHAR-GARCH1_4}).
	
	To specify the regression component $\boldsymbol{\xi}^{(j)}_t$ in equation (\ref{H-MBS-GARCH}), we define the predictor vector $\boldsymbol{z}_{i,t} = (x_{i,t-1}, x_{i,t-2}, x_{i,t-3})^\top$ for $i=1,2,3$ and $j=0,1$. Thus, the regression factor $\boldsymbol{\xi}^{(j)}_t = (\xi^{(j)}_{1,t}, \xi^{(j)}_{2,t}, \xi^{(j)}_{3,t})^\top$ in equation (\ref{H-MBS-GARCH}) is computed as:
	\[
	\xi^{(j)}_{i,t} = \beta^{(j)}_{i,1} x_{i,t-1} + \beta^{(j)}_{i,2} x_{i,t-2} + \beta^{(j)}_{i,3} x_{i,t-3},
	\]
	for $j=0,1$ and $i=1,2,3$, where the coefficients $\beta^{(j)}_{i,p}$, $p=1,2,3$, are listed in Table \ref{table2}. To evaluate the variable selection capability of Algorithm \ref{alg:SH-MBS-GARCH}, we set certain $\beta^{(j)}_{i,p}$ (for $i,p=1,2,3$ and $j=0,1$) to zero, allowing us to test whether the algorithm can accurately identify relevant explanatory variables.
	
	\begin{figure}[t!]
		\begin{center}
			\includegraphics[scale=0.75]{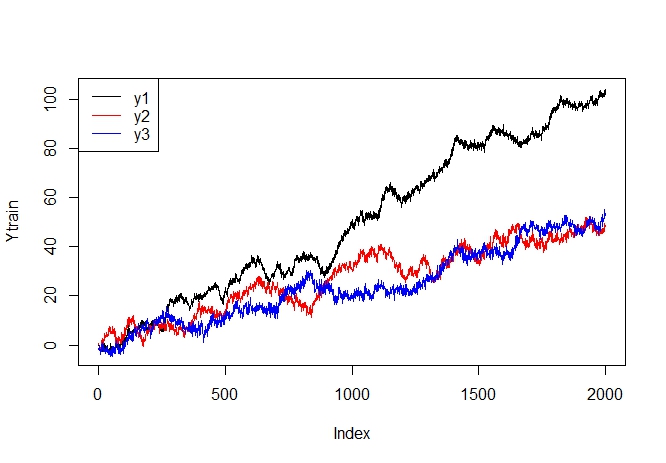}
			\caption{Time series generated from the MBS model}
			\label{fig03}
		\end{center}
	\end{figure}

	\begin{table}[!h]
		\begin{center}
			\begin{tabular}{crc|crc|crc}
				Para. & True & \shortstack{Mean \,(std.)}
				& Para. & True & \shortstack{Mean\,(std.)}
				& Para. & True & \shortstack{Mean\,(std.)} \\
				\hline
				$\tau_{1,L}$  & -0.4 & -0.46 (0.04)
				& $\tau_{2,L}$  & -0.4 & -0.79 (0.15F)
				& $\tau_{3,L}$  & -0.4 & -0.49 (0.05) \\
				$\tau_{1,U}$  &  0.4 &  0.36 (0.05)
				& $\tau_{2,U}$  &  0.4 &  0.66 (0.18)
				& $\tau_{3,U}$  &  0.4 &  0.47 (0.07) \\
				\hline
				$\sigma^2_u$  & 0.02 & 0.035 (0.011)
				& $\sigma^2_\nu$ & 0.05 & 0.069 (0.030)
				& $\sigma^2_w$  & 0.02 & 0.037 (0.0012) \\
				\hline
				$\beta^{(0)}_{1,1}$ & 2    & 1.95 (0.01)
				& $\beta^{(0)}_{2,1}$ & 2    & 1.77 (0.03)
				& $\beta^{(0)}_{3,1}$ & 2    & 1.79 (0.03) \\
				$\beta^{(0)}_{1,2}$ & 0    & -          
				& $\beta^{(0)}_{2,2}$ & 0    & -          
				& $\beta^{(0)}_{3,2}$ & 0    & -           \\
				$\beta^{(0)}_{1,3}$ & 1.5  & 1.50 (0.01)
				& $\beta^{(0)}_{2,3}$ & 1.5  & 1.66 (0.03)
				& $\beta^{(0)}_{3,3}$ & 1.5  & 1.77 (0.15) \\
				\hline
				$\beta^{(1)}_{1,1}$ & -1.5 & -1.52 (0.01)
				& $\beta^{(1)}_{2,1}$ & -1.5 & -1.55 (0.002)
				& $\beta^{(1)}_{3,1}$ & -1.5 & -1.38 (0.08) \\
				$\beta^{(1)}_{1,2}$ & 4    & 3.93 (0.01)
				& $\beta^{(1)}_{2,2}$ & 4    & 4.01 (0.002)
				& $\beta^{(1)}_{3,2}$ & 4    & 3.96 (0.10) \\
				$\beta^{(1)}_{1,3}$ & 0    & -          
				& $\beta^{(1)}_{2,3}$ & 0    & -          
				& $\beta^{(1)}_{3,3}$ & 0    & -           \\
				\hline
				$\sigma^2_\varepsilon$ & 0.55 & 0.68 (0.071)
				& -     & -     & -
				& -     & -     & -       \\
				\hline
			\end{tabular}
		\end{center}
		\caption{Means and standard deviations of the proposed estimators for $\tilde{\boldsymbol\theta_1} = (\tau_{1,L},\tau_{2,L},\tau_{3,L},\tau_{1,U},\tau_{2,U},\tau_{3,U})^\top$, $\boldsymbol{\theta}_3=(\sigma^2_u ,\sigma^2_\nu, \sigma^2_w)^\top$ and $\boldsymbol{\theta}_4=(\boldsymbol{\beta}, \sigma^2_\varepsilon)^\top$, based on 100 independent replications in the simulation study.}
		\label{table2}
	\end{table}

	\begin{table}[!h]
		\begin{center}
			\begin{tabular}{c c r | c c r | c c r}
				{Para.} & {True} & {Mean (std.)} & 
				{Para.} & {True} & {Mean (std.)} & 
				{Para.} & {True} & {Mean (std.)} \\ 
				\hline
				$\rho_1$ & 0.6   & 0.61 (0.06)  & $\rho_2$ & 0.5   & 0.52 (0.06)  & $\rho_3$ & 0.5   & 0.51 (0.06)  \\
				$s_1$    & 4     & 4 (2.49)     & $s_2$    & 4     & 4 (2.49)     & $s_3$    & 4     & 4 (2.49)     \\
				\hline
			\end{tabular}
		\end{center}
		\caption{Hyperparameter settings for $\boldsymbol{\theta}_2 = (\boldsymbol{\mu}_t, \boldsymbol{\kappa}_t)$ in the simulation study, along with the means and standard deviations of the proposed estimator, based on 100 independent replications, each using 1500 MCMC iterations with a 500-iteration burn-in period.}
		\label{table3}
	\end{table} 
	
	\begin{figure}[!h]
		\centering
		\includegraphics[scale=0.7]{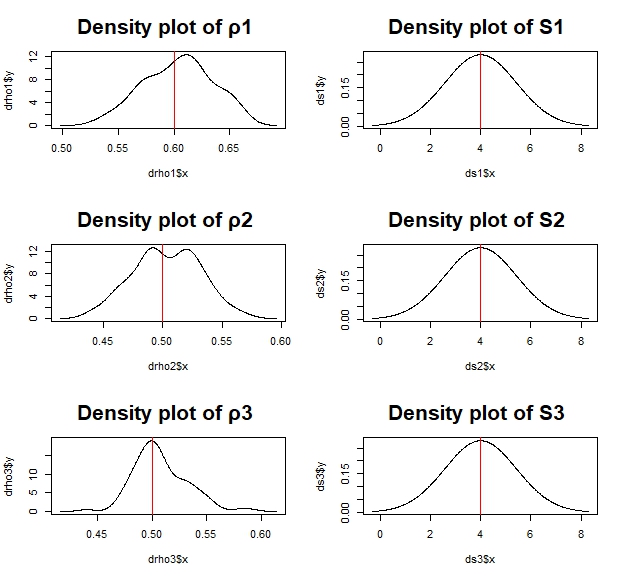}
		\caption{Density plots of the hyperparameters in Table \ref{table3}, based on 100 independent replications, with red lines indicating the true parameter values.}
		\label{fig04}
	\end{figure}

	For simplicity, we assume $\Sigma_u = \sigma^2_u I_3$, $\Sigma_v = \sigma^2_v I_3$, $\Sigma_w = \sigma^2_w I_3$, and $\Sigma_\varepsilon = \sigma^2_\varepsilon I_3$, where $I_3$ is the $3 \times 3$ identity matrix. Consequently, the parameter vectors $\boldsymbol{\theta}_3$ and $\boldsymbol{\theta}_4$ reduce to $\boldsymbol{\theta}_3 = (\sigma^2_u, \sigma^2_v, \sigma^2_w)^\top$ and $\boldsymbol{\theta}_4 = (\boldsymbol{\beta}, \sigma^2_\varepsilon)^\top$. Using the generated regime $R_t$ and regression factors $\boldsymbol{\xi}^{(j)}_t$, we simulate a three-dimensional time series $\boldsymbol{y}_t = (y_{1,t}, y_{2,t}, y_{3,t})^\top$ for $t=1,\ldots,500$ according to equation (\ref{H-MBS-GARCH}). The hyperparameter settings are reported in Table \ref{table3}. Figure \ref{fig03} displays an example of the generated $y_{i,t}$ for $i=1,2,3$, exhibiting nonstationary dynamics.

	Using the generated  time series $\boldsymbol{y}_t$ and $\boldsymbol{x}_t$ for $t=1,\ldots,500$, we apply Algorithm \ref{alg:SH-MBS-GARCH} to estimate $\boldsymbol{\theta}$. The number of explanatory variables $k$ for $\boldsymbol{\xi}^{(j)}_t$ in equation (\ref{H-MBS-GARCH}), for $j=0,1$, is treated as unknown. We set the maximum number of candidate explanatory variables to $3$. Specifically, for each regime, we consider three candidate explanatory variables $x_{i,t-p}$, $p=1,2,3$, for the $i$th time series and evaluate whether Algorithm \ref{alg:SH-MBS-GARCH} can identify the true relevant explanatory variables. For instance, only $x_{1,t-1}$ and $x_{1,t-3}$ are active in $\xi_{1,t}^{(0)}$.
	
	Based on 100 independent replications, Table \ref{table2} presents the estimated means and standard deviations of $\tilde{ \boldsymbol{\theta}}_1 = (\tau_{1,L}, \tau_{2,L}, \tau_{3,L}, \tau_{1,U}, \tau_{2,U}, \tau_{3,U})^\top$, $\boldsymbol{\theta}_3 = (\sigma^2_u, \sigma^2_v, \sigma^2_w)^\top$, and $\boldsymbol{\theta}_4 = (\boldsymbol{\beta}, \sigma^2_\varepsilon)^\top$ for $j=0,1$ and $i,p=1,2,3$. In this context,  $\theta_1$ is defined as the $q^k_j$-th quantile of the distribution of $\tilde{\theta}_1$.
	The results show that Algorithm \ref{alg:SH-MBS-GARCH} correctly identifies the true set of relevant explanatory variables in all 100 replications (100\% accuracy). Furthermore, Algorithm \ref{alg1} yields accurate and stable estimates for the corresponding regression coefficients.
	
	Additionally, we propose a procedure to estimate the hyperparameters $\rho_i$ and $s_i$ in the MBS model for the $i$th time series, $i=1,\ldots,m$, using $\boldsymbol{\delta}_t$ and $\boldsymbol{\kappa}_t$. First, the hyperparameters $\boldsymbol{D}$ and $\boldsymbol{\rho}$ in the trend component of equation (\ref{MBS}) are estimated by fitting an AR(1) model to each time series in $\boldsymbol{\delta}_t$. Second, the hyperparameter $s_i$ in the seasonal component $\boldsymbol{\kappa}_t$ is estimated through spectral analysis, utilizing the R function \texttt{spec.pgram}. Table \ref{table3} reports the estimation results for these hyperparameters, and Figure \ref{fig04} presents their posterior densities distributions, with red vertical lines marking the true values used in the simulation study.

	\subsection{Data description}
	\label{sec:Data}
	\begin{figure}[!h]
		\begin{center}
			\includegraphics[scale=0.5]{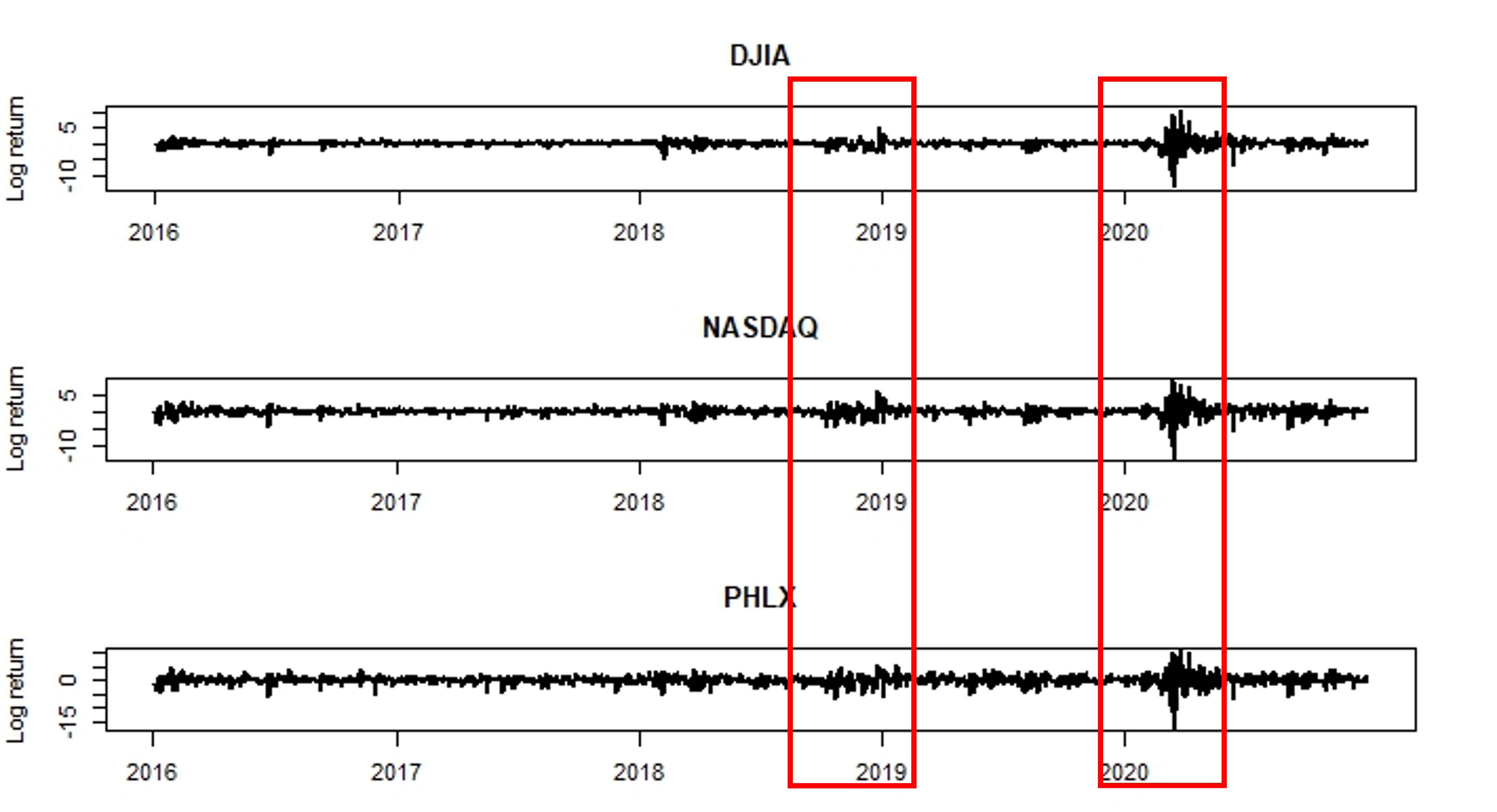}
			\caption{Daily log returns of the DJIA, NASDAQ, and PHLX indices from January 2016 to December 2020.}
			\label{fig2}
		\end{center}
	\end{figure}
	
	\begin{figure}[!h]
		\begin{center}
			\includegraphics[width=15.5cm]{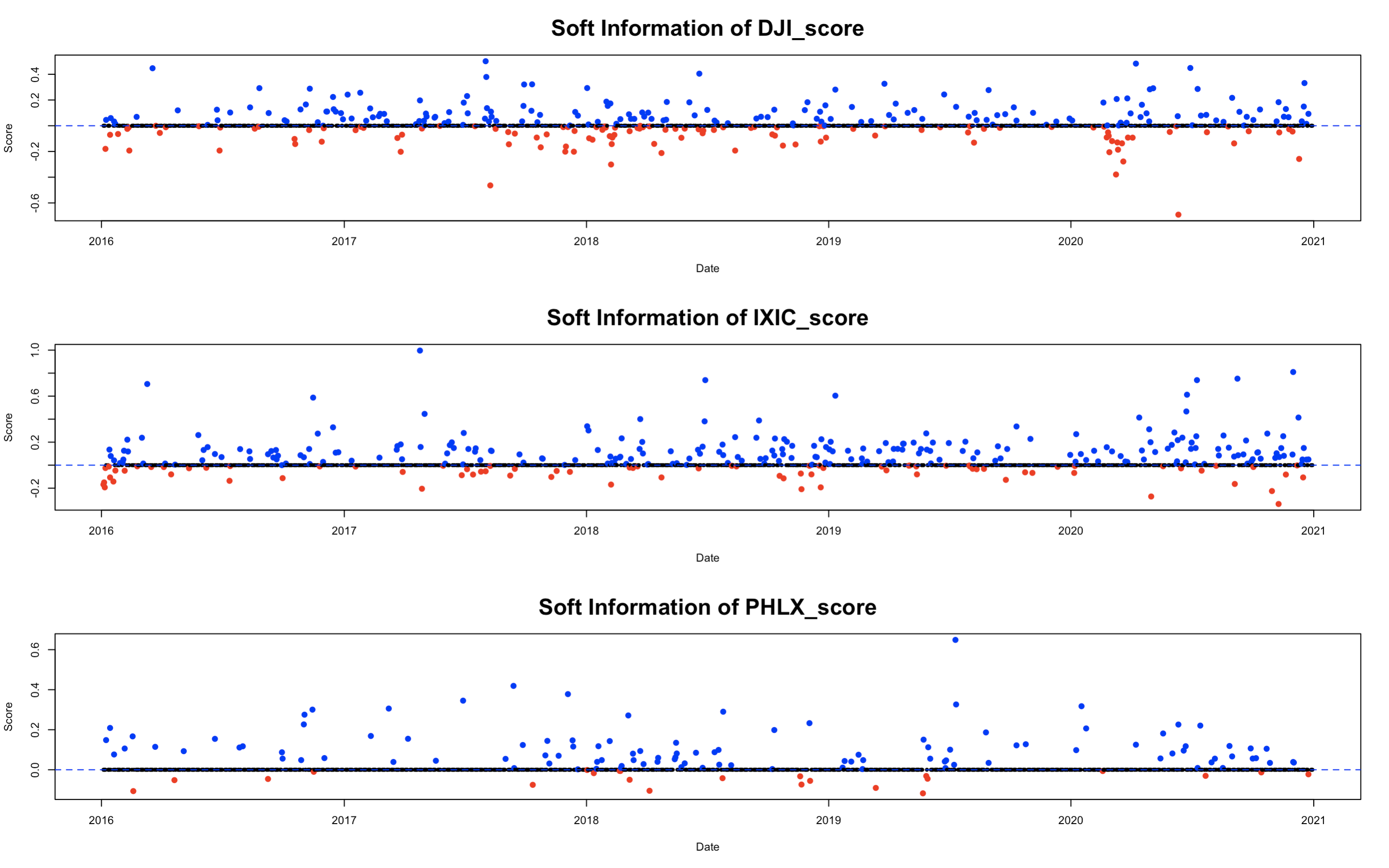}
			\caption{Sentiment scores of daily financial news from The New York Times of the DJIA, NASDAQ, and PHLX indices from January 2016 to December 2020, where blue and red points represent positive and negative scores, respectively.}
			\label{fig6}
		\end{center}
	\end{figure}

	We analyze three U.S. market indices, the Dow Jones Industrial Average (DJIA), NASDAQ Composite, and PHLX Semiconductor Index, to evaluate the empirical performance of the proposed method. Figure \ref{fig2} displays the daily log returns of these indices from January 2016 to December 2020, revealing similar return patterns across the three. Notably, two periods of high volatility, marked by red boxes, correspond to the China–U.S. trade war and the COVID-19 pandemic, including the U.S. stock market circuit breaker events. When one index exhibits high volatility, the others show similar fluctuations, suggesting that joint modeling of the three indices is preferable to individual modeling. Additionally, we extract soft information from the U.S. Daily News EPU index and daily financial news reported in The New York Times for the three indices, using the method described in Section \ref{sec:Soft}, to define the regime indicator $R_{i,t}$ in equations (\ref{SHAR-GARCH1_4a})–(\ref{SHAR-GARCH3_4a}). Figure \ref{fig6} illustrates the sentiment scores of daily financial news from The New York Times for the three indices during this period, highlighting distinct soft information dynamics for each index. Thus, the EPU-based soft information reflects broader U.S. economic conditions, while the New York Times-based soft information captures the specific impact of global economic events on each index.
	
	Let the DJIA, NASDAQ, and PHLX indices represent the 1st, 2nd, and 3rd time series, respectively. We define the explanatory vector for hard information as $\boldsymbol{z}_{i,t} = (\tilde{\boldsymbol{r}}_{t-1}^\top, \ldots, \tilde{\boldsymbol{r}}_{t-5}^\top)^\top$ for $i=1,2,3$, where $\tilde{\boldsymbol{r}}_t = (\tilde{r}_{1,t}, \tilde{r}_{2,t}, \tilde{r}_{3,t})^\top$. Thus, a Vector Autoregressive model of order 5, denoted by VAR(5), is used to construct the regression factors $\boldsymbol{\xi}^{(0)}_t$ and $\boldsymbol{\xi}^{(1)}_t$ in equation (\ref{H-MBS-GARCH}).

	\subsection{Empirical study}
	\label{sec:Empirical}
	\begin{figure}[!h]
		\begin{center}
			\includegraphics[scale=0.4]{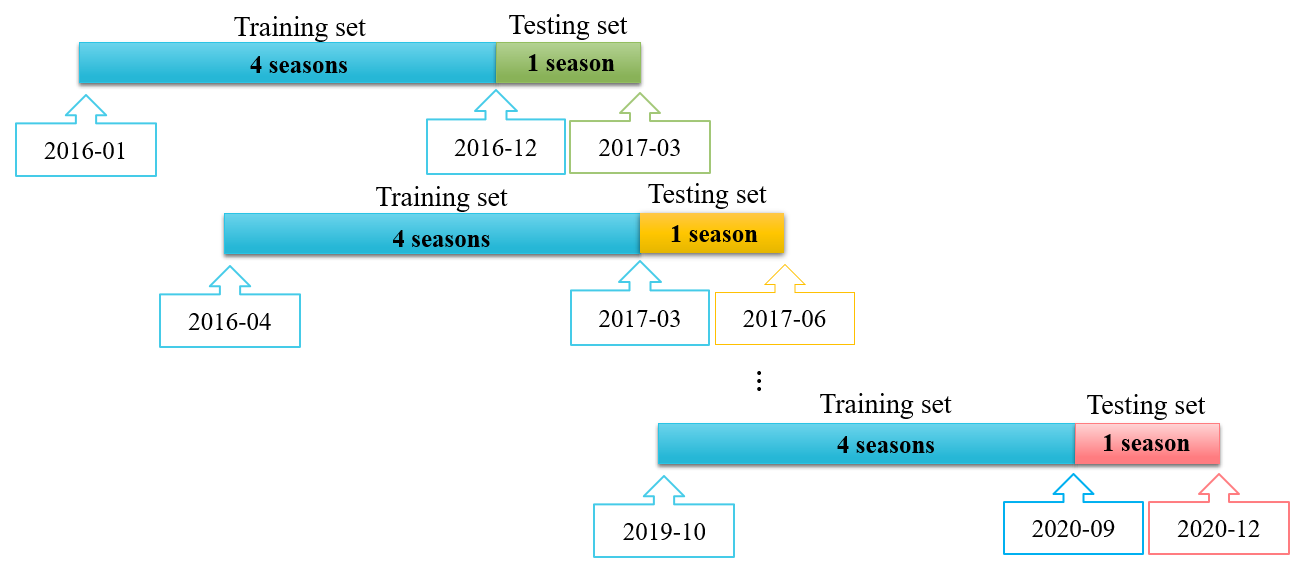}
			\caption{Flowchart of the rolling window framework from January 2016 to December 2020}
			\label{fig3}
		\end{center}
	\end{figure}

	In our empirical study, we implement a rolling window framework with a 1-year window size (252 trading days) and a quarterly update frequency (63 days). Specifically, we train the model using 1-year data, generate one-step-ahead predictions for each trading day in the subsequent quarter using fixed model parameters and updated data, and re-estimate the model at the end of the following quarter. Figure \ref{fig3} illustrates the rolling window framework flowchart from January 2016 to December 2020, resulting in 16 prediction quarters.
	
	We use mean-square error (MSE) and mean-square prediction error (MSPE) to assess the fitting and prediction performance of different models, respectively. These metrics are defined as:
	\begin{eqnarray}
		\text{MSE} &=& \frac{1}{T} \sum_{t=1}^T (y_{i,t} - \hat{y}_{i,t})^2,
		\label{mse} \\
		\text{MSPE} &=& \frac{1}{n} \sum_{t=T+1}^{T+n} [y_{i,t} - \hat{y}_{i,t-1}(1)]^2,
		\label{mspe}
	\end{eqnarray}
	where $(y_{i,1}, \ldots,y_{i,T})$ represents $T$ training samples, $\hat{y}_{i,t}$ is the fitted value for $y_{i,t}$, $(y_{i,T+1}, \ldots,$ $y_{i,T+n})$ denotes $n$ test samples, and $\hat{y}_{i,t-1}(1)$ is the one-step-ahead prediction of $y_{i,t}$ for the $i$th asset. In the rolling window framework, we set $T=252$ in equation (\ref{mse}) and $n=63$ in equation (\ref{mspe}).
	
	Table \ref{table4} reports the MSEs and MSPEs for various models across the three indices. The results demonstrate that multivariate models significantly enhance both in-sample fitting and out-of-sample prediction compared to the individual ARMA-GARCH model, indicating that the three indices share informative signals that support joint modeling and forecasting. Moreover, a comparison between the MBS-GARCH and SH-MBS-GARCH Type I models reveals that incorporating the hysteretic factor leads to improved performance.  Furthermore, the proposed SH-MBS-GARCH Type II, III, and IV models, which integrate soft information, consistently outperform or perform comparably to competing models. These empirical results underscore the benefits of utilizing multivariate structures, hysteretic dynamics, and soft information to improve modeling and predictive accuracy for the three indices.
	
	\begin{table}[t!]
		\centering
		\begin{tabular}{ccccccc}
			&ARMA- & MBS-  &\multicolumn{4}{c}{SH-MBS-GARCH} \\ \cline{4-7}
			&GARCH & GARCH &Type I  &Type II &Type III &Type IV\\ \hline
			$\exp\{$MSE$\times10^3\}$&&&&&&\\ 
			DJIA   
			&3.380 &1.165 &1.147 & 1.138  & {\bf1.136}&{\bf1.144} \\
			NASDAQ &3.314 &1.149 &1.152 &1.126 &{\bf1.120} &{\bf1.116} \\
			PHLX   &1.802 &1.309 &1.301 &1.271 & {\bf1.261} &{\bf1.264} \\ \hline
			$\exp\{$MSPE$\times10^3\}$&&&&&&\\
			DJIA   &1.858 &1.551 &1.486 &{\bf1.427} &1.437 & {\bf1.410} \\
			NASDAQ &1.727 &1.506 &1.442 & {\bf1.360} &{\bf1.382} &1.412 \\
			PHLX   &2.055 &1.899 &1.763 &1.714 &{\bf1.709} & {\bf 1.670} \\ \hline
		\end{tabular}
		\caption{Average $\exp\{\text{MSE} \times 10^3\}$ and $\exp\{\text{MSPE} \times 10^3\}$ across 16 prediction quarters for various models applied to the three indices, with the two smallest values in each row highlighted in bold.}
		\label{table4}
	\end{table}
	
\begin{figure}[t!]
		\begin{center}
			\includegraphics[scale=0.7]{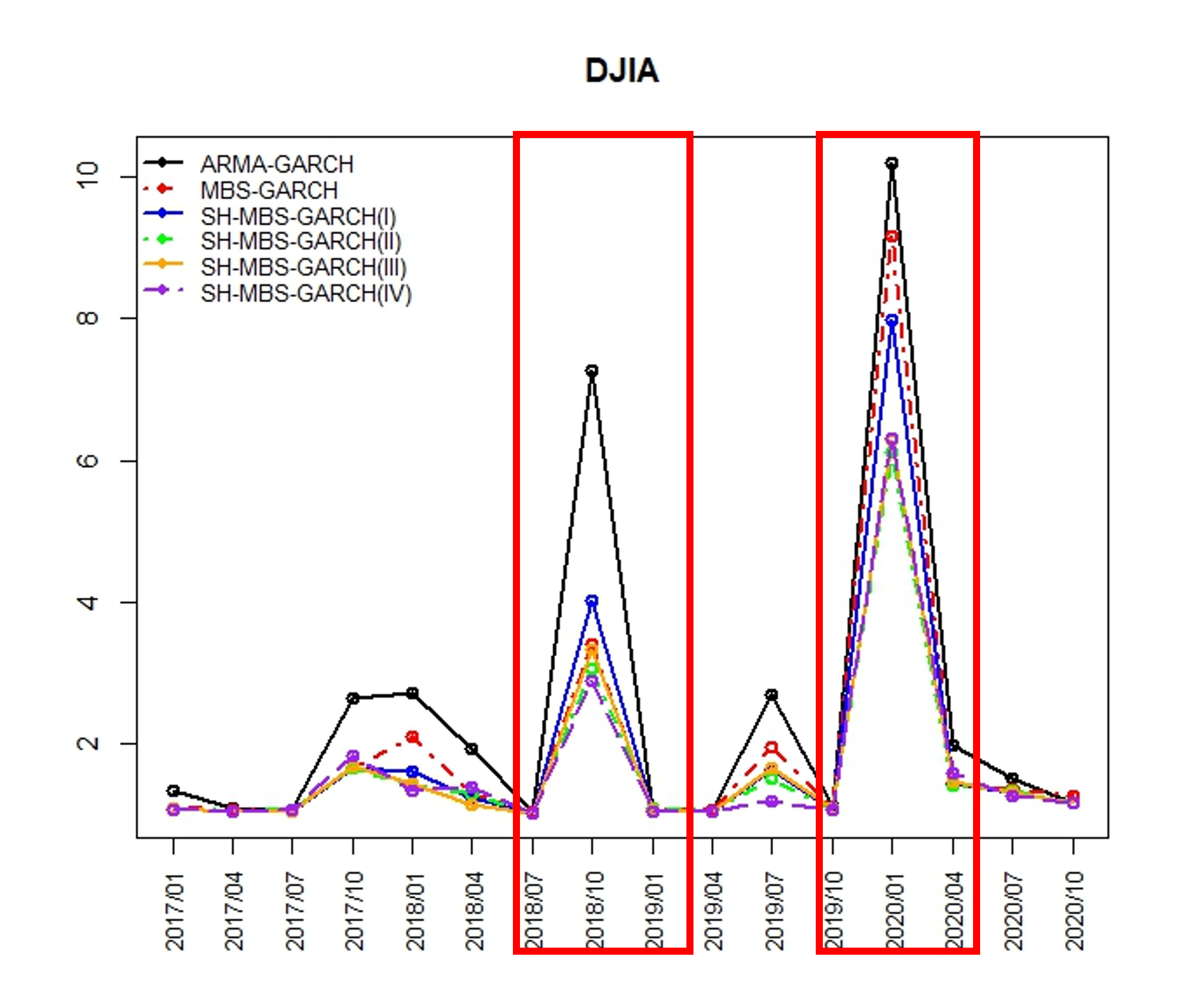}
			\caption{The $\exp\{\text{MSPE} \times 10^3\}$ for the ARMA-GARCH (black), MBS-GARCH (red), SH-MBS-GARCH Type I (blue), Type II (green), Type III (orange), and Type IV (purple) models for the DJIA over the study period.} 
			\label{fig8}	   
		\end{center}
	\end{figure}

	Figure \ref{fig8} displays the MSPEs for the DJIA across the ARMA-GARCH (black), MBS-GARCH (red), and SH-MBS-GARCH Type I (blue), Type II (green), Type III (orange), and Type IV (purple) models during the study period. Periods associated with the China– United States trade war and the COVID-19 pandemic, during which circuit breakers halted U.S. stock trading, are highlighted in red boxes, corresponding to the periods identified in Figure \ref{fig2}. The NASDAQ and PHLX indices exhibit similar patterns and are omitted for brevity. Aligning with Table \ref{table4}, the SH-MBS-GARCH models achieve lower MSPEs than competing models across most of the 16 quarters. Notably, the SH-MBS-GARCH Type IV model, which uses only soft information to define regime indicators, exhibits superior predictive performance during high-volatility periods. This result underscores the importance of incorporating soft information for enhancing market trend prediction, especially during periods of elevated risk.
	
	\section{Conclusion}
	\label{sec:conclusion}
	This study introduces the SH-MBS-GARCH model to capture multivariate time series dynamics, effectively integrating soft information in novel ways. A Bayesian algorithm, combining adaptive MCMC and spike-and-slab regression, is developed to estimate model parameters, with simulation results confirming that Algorithm \ref{alg:SH-MBS-GARCH} delivers accurate estimates for threshold and regression parameters, alongside straightforward hyperparameter estimation. Empirical analysis of the DJIA, NASDAQ, and PHLX indices demonstrates superior fitting and prediction accuracy compared to competing models. While this paper focuses on traditional time series forecasting targeting the mean of the joint forecast distribution, joint quantile forecasting offers significant potential for economic applications, such as value-at-risk estimation in finance for trading and hedging strategies. Recently, \cite{ning2025bayesian} explored Bayesian feature selection for joint quantile time series analysis within the MBS framework, and we propose extending regime-switching models in this direction for future research.

	\section*{Declaration of Interests}
	Ning Ning serves as an Associate Editor for the Journal of Computational and Graphical Statistics.
	
	\appendix
	\renewcommand{\thesection}{Appendix \Alph{section}}
	\setcounter{section}{1}  
	\section*{Appendix}
	
	\subsection{Adaptive MCMC}
	\label{secA.1}
	
	Direct sampling from the posterior distribution is impractical due to the non-standard form of the conditional posterior distributions. To overcome this, we utilize the random walk Metropolis (RWM) algorithm for generating MCMC estimates \citep{Mh1953,Hasting1970}. In cases where the RWM algorithm shows high correlations or slow convergence, we implement the adaptive MCMC sampling approach of \cite{CS2006}, which employs RWM during the burn-in phase before switching to the independent kernel Metropolis-Hastings algorithm to enhance convergence and mixing. For clarity, we group the parameters of interest into specific sets. The adaptive MCMC sampling method is detailed below.
	\begin{algorithm}
		\caption{Adaptive MCMC Sampling for Parameter Estimation}\label{alg:adaptive_mcmc}
		\begin{algorithmic}[1]
			\State Initialize the parameter vector $\boldsymbol{\theta}^{(0)}$ and set $i = 0$.
			
			\While{$i < N$} 
			\If{$i < M$} \Comment{$M$ is the burn-in period}
			\State Generate candidate $\boldsymbol{\theta}^{(i)} \sim \mathcal{N}(\boldsymbol{\theta}^{(i-1)}, c \Sigma)$, where $c$ is a scaling constant and $\Sigma$ is \\
			\hspace{3cm} the covariance matrix.
			\State Accept $\boldsymbol{\theta}^{(i)}$ with probability $\alpha = \min\left(1, \frac{p(\boldsymbol{\theta}^{(i)})}{p(\boldsymbol{\theta}^{(i-1)})}\right)$, where $p(\cdot)$ is the conditional \\
			\hspace{3cm} posterior distribution. If not accepted, set $\boldsymbol{\theta}^{(i)} = \boldsymbol{\theta}^{(i-1)}$.
			\Else
			\State Generate candidate $\boldsymbol{\theta}^{(i)} \sim \mathcal{N}(\bar{\boldsymbol{\theta}}, \Sigma_{\text{adapt}})$, where $\bar{\boldsymbol{\theta}}$ and $\Sigma_{\text{adapt}}$ are the mean and \\
			\hspace{3cm} covariance matrix estimated from burn-in iterates.
			\State Accept $\boldsymbol{\theta}^{(i)}$ with probability 
			\[
			\alpha = \min\left(1, \frac{p(\boldsymbol{\theta}^{(i)}) f(\boldsymbol{\theta}^{(i-1)} \mid \bar{\boldsymbol{\theta}}, \Sigma_{\text{adapt}})}{p(\boldsymbol{\theta}^{(i-1)}) f(\boldsymbol{\theta}^{(i)} \mid \bar{\boldsymbol{\theta}}, \Sigma_{\text{adapt}})}\right),
			\]
			\hspace{3cm} where $f(\cdot \mid \bar{\boldsymbol{\theta}}, \Sigma_{\text{adapt}})$ is the Gaussian proposal density. If not accepted, \\
			\hspace{3cm} set $\boldsymbol{\theta}^{(i)} = \boldsymbol{\theta}^{(i-1)}$.
			\EndIf
			\State Increment $i \gets i + 1$.
			\EndWhile
		\end{algorithmic}
	\end{algorithm}

	\subsection{Spike-and-slab regression}
	\label{secA.2}
	
	Spike-and-slab regression is a widely used Bayesian variable selection method, renowned for its ability to effectively capture sparsity in model structures \citep{GM1993, Lai2025}. In high-dimensional regression settings, where many regression coefficients are expected to be zero, the spike-and-slab model employs a mixture prior for each coefficient: a ``spike" component, tightly concentrated around zero for irrelevant variables, and a ``slab" component with larger variance for potentially relevant variables, enabling robust and flexible variable selection. This approach eliminates the need to pre-specify a fixed set of explanatory variables, as the relevance of each predictor is dynamically updated through posterior inference based on observed data, allowing estimation of inclusion probabilities while maintaining model flexibility and accounting for uncertainty.

	\subsection{The simulation smoother method}
	\label{secA.4}
	
	\cite{DK2002} considered the following state space model:
	\begin{equation}
		\begin{aligned}
			\boldsymbol{y}_t &= Z_t \boldsymbol{\alpha}_t + \boldsymbol{\varepsilon}_t, &\quad \boldsymbol{\varepsilon}_t &\sim \mathcal{N}(0, H), \\
			\boldsymbol{\alpha}_{t+1} &= T_t \boldsymbol{\alpha}_t + R_t \boldsymbol{\eta}_t, &\quad \boldsymbol{\eta}_t &\sim \mathcal{N}(0, Q),
		\end{aligned}
		\label{secA.4.1}
	\end{equation}
	where $\boldsymbol{y}_t$ is a $p \times 1$ observation vector, $\boldsymbol{\alpha}_t$ is an $m \times 1$ state vector, and $\boldsymbol{\varepsilon}_t$ and $\boldsymbol{\eta}_t$ are disturbance vectors for $t = 1, \ldots, n$. The matrices $Z_t$, $T_t$, $R_t$, $H$, and $Q$ are assumed known, with the initial state $\boldsymbol{\alpha}_1 \sim \mathcal{N}(\boldsymbol{a}_1, P_1)$, where $\boldsymbol{a}_1$ and $P_1$ are known. We later explore cases where components of $\boldsymbol{a}_1$ and $P_1$ are unknown. In the context of the SH-MBS-GARCH model, the notations in equation (\ref{secA.4.1}) are specified as follows: $\boldsymbol{y}_t = \boldsymbol{Y} - \boldsymbol{\xi}^{(j)}_t$, $\boldsymbol{\alpha}_t = (\boldsymbol{\mu}_t, \boldsymbol{\kappa}_t)^{\top}$, $H = \Sigma_\varepsilon$, $\boldsymbol{\eta}_t = (\boldsymbol{u}_t, \boldsymbol{v}_t, \boldsymbol{\omega}_t)^{\top}$, and $Q = \text{diag}(\Sigma_u, \Sigma_v, \Sigma_w)$, where the right-hand side notations are defined in equation (\ref{H-MBS-GARCH}).
	
	Following \cite{DK2002}, the simulation smoother procedure begins by generating an unconditional latent state path and its corresponding observation path from the prior structure of the state space model, independent of observed data. Specifically, we draw the initial state from the prior distribution as $\alpha_1^+ \sim \mathcal{N}(\boldsymbol{a}_1, P_1)$. For each time point $t = 1, \ldots, n$, we then simulate state disturbances $\eta_t^+ \sim \mathcal{N}(0, Q)$ and observation disturbances $\varepsilon_t^+ \sim \mathcal{N}(0, H)$, updating the latent state via the transition equation $\alpha_{t+1}^+ = T_t \alpha_t^+ + R_t \eta_t^+$ and generating the observation as $y_t^+ = Z_t \alpha_t^+ + \varepsilon_t^+$. This yields simulated samples from the joint prior distribution $p(\alpha_{1:n}, y_{1:n})$.
	
	Next, we apply the Kalman filter and smoother independently to the actual observations $\boldsymbol{y}$ and the simulated data $\boldsymbol{y}^+$ to compute smoothed state expectations, denoted as $\hat{\boldsymbol{\alpha}}_t$ for the real data and $\hat{\boldsymbol{\alpha}}_t^+$ for the simulated data, using the forward and backward recursions described by \cite{DK2002}. Subsequently, we generate a posterior sample $\tilde{\boldsymbol{\alpha}}_t$ from the conditional distribution $p(\boldsymbol{\alpha} \mid \boldsymbol{y})$ by adjusting the unconditional simulated path with the difference between the smoothed means, defined as $\tilde{\boldsymbol{\alpha}}_t = \hat{\boldsymbol{\alpha}}_t + (\boldsymbol{\alpha}_t^+ - \hat{\boldsymbol{\alpha}}_t^+)$, thereby incorporating both prior randomness and information from the observed data to produce a correctly distributed posterior sample.

	
	\bibliography{SH-MBS-GARCH-bib}
	
\end{document}